\documentclass{aa}  
\usepackage{graphicx,color}
\usepackage{txfonts}
\usepackage{comment}
\usepackage{amsmath}
\usepackage{xcolor}
\usepackage{caption, multirow}
\DeclareCaptionLabelFormat{bf}{\textbf{#1 #2}}
\usepackage{natbib,twoopt}
\usepackage[hyphenbreaks]{breakurl}
\usepackage[breaklinks]{hyperref}  
\DeclareMathAlphabet\mathzapf       {T1}{pzc} {mb} {it}
\bibpunct{(}{)}{;}{a}{}{,}    

\definecolor{cobalt}{rgb}{0.06, 0.2, 0.65}
\hypersetup{
  colorlinks,
  citecolor=cobalt,
  linkcolor=cobalt,
  urlcolor=cobalt,
}
\makeatletter
  \newcommandtwoopt{\citeads}[3][][]{\href{http://adsabs.harvard.edu/abs/#3}%
    {\def\hyper@linkstart##1##2{}%
     \let\hyper@linkend\@empty\citealp[#1][#2]{#3}}}
  \newcommandtwoopt{\citepads}[3][][]{\href{http://adsabs.harvard.edu/abs/#3}%
    {\def\hyper@linkstart##1##2{}%
     \let\hyper@linkend\@empty\citep[#1][#2]{#3}}}
  \newcommandtwoopt{\citetads}[3][][]{\href{http://adsabs.harvard.edu/abs/#3}%
    {\def\hyper@linkstart##1##2{}%
     \let\hyper@linkend\@empty\citet[#1][#2]{#3}}}
  \newcommandtwoopt{\citeyearads}[3][][]%
    {\href{http://adsabs.harvard.edu/abs/#3}
    {\def\hyper@linkstart##1##2{}%
     \let\hyper@linkend\@empty\citeyear[#1][#2]{#3}}}
\makeatother

\begin{document}

\title{Search for the multiwavelength counterparts to extragalactic unassociated \textit{Fermi} $\gamma$-ray sources}
\authorrunning{Ulgiati et al.}


   \author{A. Ulgiati
          \inst{1,2}, S. Paiano \inst{1}, F. Pintore \inst{1}, T. D. Russell \inst{1}, B. Sbarufatti \inst{3}, C. Pinto \inst{1}, E. Ambrosi \inst{1}, A. D'Aì \inst{1}, G. Cusumano \inst{1}, M. Del Santo \inst{1}
         }

   \institute{INAF, Istituto di Astrofisica Spaziale e Fisica Cosmica, Via U. La Malfa 153, I-90146 Palermo, Italy \\
              \email{alberto.ulgiati@inaf.it}
         \and
             Universit\`a degli Studi di Palermo, Dipartimento di Fisica e Chimica, via Archirafi 36, I-90123 Palermo, Italy
         \and    
             INAF - Osservatorio Astronomico di Brera, via Bianchi 46, I-23807, Merate (Lecco), Italy
             }

   \date{Received XXX; accepted XXX}


 \abstract
 {}
{In this paper, we searched for multi-wavelength (X-ray, optical and radio) counterparts to the unassociated gamma-ray sources (UGS) of the \textit{Fermi} 4FGL-DR4 catalog. The main goal is to identify new blazars and/or new active galactic nuclei (AGNs) emitting at GeV energies [like (Narrow Line) Seyfert-1 and radio galaxies].}
{We focus on sky regions observed by the \textit{Swift} satellite that overlap with the reported positions of the UGSs. Since our primary interest lies in extra-galactic sources, we focus on UGSs located outside the Galactic plane ($|b|>10^{\circ}$). Due to the large number of sources (about 1800 UGS), we developed a pipeline to automatise the search for counterparts and significantly reduce the computational time for the analysis. Our association process begins by identifying potential X-ray counterparts for each UGS; if one is found, we further look  for corresponding radio and optical counterparts in the X-ray counterpart error box, thus minimizing ambiguities.}
{Out of the 1284 UGSs in the 4FGL-DR4 catalog, 714 were observed at least once by \textit{Swift}/XRT. We detected, with a significance of $\geq$ 3$\sigma$, at least one X-ray source within the \textit{Fermi} error box for 274 of these $\gamma$-ray emitters. Among these, 193 UGSs have a single potential X-ray counterpart (referred to as UGS1), while 81 have multiple potential X-ray counterparts within the Fermi error box (referred to as UGS2). Of the UGS2, 54 have two X-ray counterparts, 11 have three, and the remaining 16 have more than three. Each UGS1 has an optical counterpart, and 113 also could be associated to a radio counterpart. We performed a comparison of the possible counterpart properties with those of the $\gamma$-ray emitters identified by \textit{Fermi}, with the aim to assess the goodness of our associations.}
{}


   \maketitle
%

\section{Introduction}  
\label{sec:introduction} 

Fifteen years have elapsed since the launch of the \textit{Fermi} satellite, a space observatory designed for studying celestial objects emitting in the $\gamma$-ray energy range. With a field of view of 2.4 sr, \textit{Fermi} performs an all-sky survey every three hours, being an exceptionally powerful tool for studying $\gamma$-ray sources in the 100 MeV -- 300 GeV energy range \citep[for details see][]{Atwood2009}.

In 2023, the \textit{Fermi} collaboration released the incremental version of the fourth catalog \citep[4FGL-DR4,][]{4FGL_cat,4FGL_DR4}, which reports the sources detected over the 14 years of observations by the \textit{Large Area Telescope} (LAT) detector. Of the total 7195 objects  in the catalog, 4765 have been associated or identified at other wavelengths thanks to either a positional overlap in the sky, measurements of correlated variability, and/or multi-wavelength spectral properties \citep[][]{4FGL_cat,4FGL_DR4}. The catalog includes both Galactic sources and extra-galactic sources, with the latter which dominate the $\gamma$-ray sky. 
Active Galactic Nuclei (AGN) are galaxies with a super-massive black accreting matter from the surrounding galactic regions. They are the key component of these extra-galactic sources. Among the AGN, blazars represent the most numerous population, with 3934 objects identified in the catalog. Blazars are distinguished by their jets oriented at a small angle ($\theta < 10^{\circ}$) relative to the line of sight of the observer \citep[e.g.][]{Urry_1995, Giommi_2013, Costamante_2020}. Other classes of AGN, such as radio galaxies and Seyfert galaxies, constitute only a small fraction (approximately 1\%) of the catalog \citep[e.g.][]{Cheung_2010, Abdo_2010, Ackermann_2011, Grandi_2012, Paliya_2015, Angioni_2017, Rieger_2017, Jarvela_2021, Ye_2023, 4FGL_DR4}. 

Due to this geometrical configuration, the observed blazar emission is dominated by the relativistic jet, whose emission spans from the radio band up to very high energies (VHE, $E >$ 100 GeV) $\gamma$-rays. The spectral energy distribution (SED) exhibits a double bump shape: the first bump extends from sub-mm to the X-ray energy band and is interpreted as synchrotron emission from a population of relativistic electrons in the jet, while the second bump dominates the emission in the MeV--TeV range but its origin still remains a matter of debate.
The two main scenarios to explain this second high-energy peak involve either hadronic or leptonic processes. In the leptonic scenario, the second peak is attributed to inverse Compton scattering, where the same population of relativistic electrons that produce the synchrotron emission up-scatter low-energy photons, from the jet itself and/or from external regions (disc, torus or broad line region), to higher energies \citep[][]{costomante2018}. Conversely, in the hadronic scenario the high-energy emission is due to interactions involving relativistic protons, such as proton synchrotron radiation or photo-pion production, followed by cascading processes that produce $\gamma$-rays \citep[e.g.][]{Cerruti2011, Rodrigues2019, Gao2019, Cerruti2020}. The blazar population is categorized into two major classes based on the properties of their optical spectra: flat spectrum radio quasars (FSRQs), characterized by prominent emission lines, traditionally defined as having an equivalent width, $EW  >  5 \mathring{A}\ $) \citep{Stickel_1991, Stocke_1991, falomo_2014}, and BL Lacertae objects (BL Lacs or BLL), with weak or absent emission lines ($EW  <  5 \mathring{A}\ $).

However, within the \textit{Fermi} catalog there exists a substantial population (about 30\% of sources) that remains unclassified, having no yet known association with any lower energy counterparts, hereafter referred to as unassociated gamma-ray sources (UGSs). UGSs represent a significant component of the high-energy sky, possibly hiding new blazars and/or new AGN of different classes. Moreover, since they are faint objects (on average UGSs have lower $\gamma$-ray fluxes, $\sim 5.3 \times 10^{-12}$ erg cm$^{-2}$ s$^{-1}$, with respect to associated AGN sources, $\sim 1.6 \times 10^{-11}$ erg cm$^{-2}$ s$^{-1}$, in the 100 MeV to 100 GeV range), they may represent a higher redshift AGN population and/or lower luminosity sources. Therefore, their investigation is key to population studies, the development of physical models, and the interpretation of the cosmic evolution of the $\gamma$-ray sources \citep[][]{Ajello_2014,Ghisellini_2017}. In addition, the UGSs identification is crucial for the estimate of the VHE cosmic background \citep[][]{Dimauro_2018,Acciari_2019,4FGL_DR4}.

The main difficulty when selecting UGS counterparts is due to the large \textit{Fermi} positional errors arising from limited photon statistics and the angular resolution of the LAT detector (68\% containment radius: 3.5$^{\circ}$ at 100\,MeV, 0.6$^{\circ}$ at 1\,GeV, and >0.15$^{\circ}$ at $>$10 GeV). This results in the presence of tens to hundreds of potential low-energy (optical and radio bands) counterparts within the typical \textit{Fermi} error boxes.  The number of candidates is severely affected by several factors as the energy band, the detector sensitivity and the sky region. Hence, attributing the $\gamma$-ray emission to a specific source among these candidates can therefore be very challenging.
Several efforts have been made over the years in order to associate and classify UGSs. Various statistical algorithms \citep[][]{Ackermann_2012,Mirabal_2012,Mao_2013} and neural network classifiers \citep[][]{Doert_2014,Salvetti_2017b,Kaur_2019,Kaur_2019b,Kerby_2021,Kaur_2023}, based on $\gamma$-ray spectral features and variability information, have been developed with the aim to distinguish blazars from other $\gamma$-ray source populations. Other works were focused on the search for counterparts at other wavelengths.  In the radio band \citet[][]{Petrov_2013}, \citet[][]{Nori_2014}, \citet[][]{Schinzel_2015}, \citet[][]{Schinzel_2017}, \citet[][]{Bruzewski_2021}, \citet[][]{Bruzewski_2022} provided radio counterparts of UGSs using ATCA and LOFAR data. \citet[][]{Dabrusco_2013} and \citet[][ and references therein]{Massaro_2015} proposed AGN candidates on the basis of the colours of the infrared (IR) counterparts in the Wide-field Infrared Survey Explorer (WISE) survey \citep[][]{Wright_2010}. Searches using X-ray data has been performed by \citet[][]{Falcone_2011}, \citet[][]{Takahashi_2012}, \citet[][]{Landi_2015}, \citet[][]{Salvetti_2017a}, \citet[][]{Kaur_2019}, \citet[][]{Kaur_2019b}, \citet[][]{Kerby_2021}, \citet[][]{Kaur_2023} and \citet[][]{Marchesini_2020}, while a multiwavelength approach has been adopted by \citet[][]{Acero_2013}, \citet[][]{Paiano_SED}, \citet[][]{Fronte_2023}. Moreover, numerous pulsars have been discovered among the UGSs through timing studies from the $\gamma$-ray band to the radio band \citep[][and references therein]{Abdo_2013,Wu_2018,Li_2018}.

The firm classification of extragalactic UGSs comes from studying the optical spectra of their lower energy counterparts. Optical spectra act as the fingerprints of these objects, enabling their extragalactic nature and classification to be determined. \citet[][]{Shaw_2012,Shaw_2013} studied the optical spectra of about 500 sources of the second \textit{Fermi} catalog \citep[2FGL, ][]{Nolan_2012}, confirming the BLL nature of these objects and estimating a redshift (or a lower limit) for each of them. \citet[][]{Dabrusco_2013}, \citet[][]{Massaro_2015,Massaro_2016} and references therein, focused on the association and classification of the UGSs of the second and third \textit{Fermi} \citep[3FGL,][]{Acero_2015} catalogs. Among a sample of $\sim$600 candidates, they classified $\sim$200 blazars using selection criteria based on the infrared (IR) colours of counterparts in the WISE survey. \citet{Paiano_2017a, Paiano_2017b, Paiano_2020, Paiano_2021b, Paiano_2023} analysed the optical spectra of $\sim$150 BLL objects (or candidates) identified as TeV (or TeV candidate) emitters and/or potential neutrino sources.
In addition, \citet[][]{paiano2017_ufo1,paiano2019_ufo2,Ulgiati_2024} focused on the association and classification of UGSs from the second, third and forth \textit{Fermi} \citep[4FGL, 4FGL-DR3, 4FGL-DR4][]{4FGL_cat,4FGL_DR3,4FGL_DR4} catalogs. This resulted in the classification of $\sim$80 new AGN (of which $\sim$ 80\% were blazars, and $\sim$ 20\% were AGN of other types).  

In this work we analyse UGSs from the 4FGL-DR4 catalog that are located outside of the Galactic plane ($|b|>10^{\circ}$) that were also observed by the \textit{Neil Gehrels Swift Observatory} \citep[][]{Gehrels_2004}. The choice to exclude sources in the region of the Galactic plane is motivated by our primary interest in AGN, as sources at high Galactic latitudes are more likely to be extragalactic. For more than 10 years, the \textit{Swift} satellite has been involved in a campaign dedicated to observing UGSs \citep[][]{Stroh_2013,Falcone_2014} and all data are available on the public archive\footnote{https://www.swift.ac.uk/swift\_portal/}. The aim of this paper is to find X-ray counterparts of a sample of UGSs. This allows us to reduce the degeneracy in the optical counterpart determination, as the number of X-ray sources in the typical \textit{Fermi} error boxes (the average 99.7\% containment radius is $\sim$ 6 arcminutes, estimated over the entire range of LAT) is much smaller than in the optical band or even at lower energies. Furthermore, it also narrows the search region for identifying potential radio and optical counterparts, given that the angular resolution of the \textit{Swift}/XRT detector is a few arcsec. Our search for UGS counterparts is based on the analysis of all XRT images, covering the UGS sky regions.

The paper is structured as follows: Section \ref{sec:pipeline} outlines the X-ray data reduction and analysis, automated through the use of a pipeline developed by our group; in Section \ref{sec:MWL_counterparts} the search for the multiwavelength counterparts is described; results are discussed in Section \ref{sec:results}; section \ref{sec:discussion} contains the discussion; while section \ref{sec:conclusions} contains the conclusions.

\section{X-ray detection pipeline, data reduction and analysis}
\label{sec:pipeline} 

The number of UGSs in the 4FGL-DR4 catalog is 2430, of which 1284 are located outside of the Galactic plane region. 
Out of those 1284 sources, 714 are covered by at least one \textit{Swift}/XRT observation.

Given the large number of UGSs with available X-ray data, we developed a pipeline to reduce and automatize the data reduction and analysis process. The pipeline is Python based, performing source detection and spectral analysis on all of the available XRT observations overlapping the UGS positions. The HeaSoft FTOOLS version 6.30.1 \citep[][]{Nasa_2014} package was used for analysis. Using the data reduction tool provided by the UK \textit{Swift} Science Data Centre\footnote{https://www.swift.ac.uk/user\_objects/} \citep[][]{Evans_2020}, for each UGS the pipeline takes the UGS position and performs the following:

\begin{itemize}
    \item look for all the available Swift-XRT observations that cover the UGS positions within a cone search radius of 20 arcminutes;
    \item create a 0.3--10 keV \textit{Swift}/XRT stacked image;
    \item run a source-detection algorithm (with an initial 1.5 $\sigma$ significance cutoff);
    \item determine the position (with associated uncertainty) of any detected X-ray source within the \textit{Swift}/XRT field of view (23.6' $\times$ 23.6'), around the position of the $\gamma$-ray emitter. If data acquired with the UV and Optical Telescope (UVOT) on board of Swift are available, it provides the enhanced source position corrected for astrometry \citep[][]{Goad_2007};
    \item generate a list of X-ray detections (to which an ID is assigned), listing the position and the Signal-to-Noise Ratio (SNR) for each X-ray detected source.
    \item select detected sources that are within the $3\sigma$ \textit{Fermi} error region of UGS sources with a SNR $\geq$3\footnote{The two axes of the 4FGL-DR4 error ellipses at a 95\% confidence level have been increased by 50\% in order to yield $\sim$99\% confidence level}, 
    \item for each of the selected X-ray detections, an average spectrum is produced. Circular (annular) extraction regions are used to extract the source (background) events, as described in \citet{Evans_2009};
    \item each spectrum is grouped with the HeaSoft FTOOLS tool {\sc grppha} in order to accumulate at least 8 counts per energy bin. This choice is a compromise to maximize the degrees of freedom of the fit and the significance of the points, given the small number of collected counts for each X-ray source (the median value is $\sim$ 40 photons). Additionally, this grouping allows the use of $\chi^2$ statistics for the analysis for most of the analysed spectra\footnote{Simulations were performed to assess the reliability of $\chi^2$ statistics for this sample, and the results indicate that the statistics are equally reliable when using 8 counts per bin as when using 25 counts per bin.},
    \item finally, the 0.3 -- 10 keV X-ray spectral analysis is performed using {\sc xspec} version 12.12.1 \citep[][]{Arnaud_1996,Arnaud_2022}. 
\end{itemize}

Since blazars are the dominant population in the 4FGL-DR4 catalog and are characterised by a non-thermal X-ray emission, we model the spectra of each detected X-ray source with an absorbed power law. In particular, we fit the data with the model {\sc tbabs*powerlaw} in {\sc xspec}, where {\sc tbabs} describes the column density (nH) of the interstellar and intergalactic absorption, adopting the abundance set from \citet{Wilms_2000}.
Given the small number of detected counts per source, the \textit{nH} parameter was fixed to the Galactic line-of-sight value. Furthermore, in order to have a rough estimate of the source flux, we note that the powerlaw photon index was fixed to 2 for spectra that had no more than 2 or 3 spectral points. The photon index of 2 was determined as a reasonable average of the typical blazar slope \citep[e.g.]{Padovani_2017}.

\section{Search for multiwavelength counterparts}
\label{sec:MWL_counterparts}

Taking the position of X-ray sources found within the UGS error ellipses, we search for their possible radio and optical counterparts. To do so, we take the $\sim$99\% confidence level error boxes on the X-ray positions (which is on average $\sim$ 4 arcsec).

In the optical, we search for counterparts using the Sloan Digital Sky Survey \citep[SDSS, ][]{Ahumada_2020}, the Panoramic Survey Telescope and Rapid Response System \citep[PanSTARRS, ][]{Chambers_2016} database, the Dark Energy Survey \citep[DES, ][]{Abbott_2021} and the SuperCOSMOS Sky Survey \citep[SSS, ][]{Hambly_2004}. 
Once possible optical counterparts have been identified, we then searched the \textit{GAIA} Early Data Release 3 \citep[EDR3, ][]{Seabroke_2021} archive to determine which sources have measured proper motions. The significance of the proper motion is calculated as the ratio between the proper motion and its uncertainty. 

Two catalogs were used to search for possible radio counterparts: the Very Large Array Sky Survey \citep[VLASS;][]{Lacy_2020}, and the Rapid ASKAP Continuum Survey \citep[RACS;][]{Hale_2021}.  
If a radio counterpart is not significantly detected/reported in the radio catalogs but was within the survey region, we set an upper limit flux using the sensitivity threshold of the RACS survey (1.5 mJy at 5$\sigma$) for the sources located at -90$^{\circ}$ < $\delta$ < -40$^{\circ}$ and of VLASS survey (0.345 mJy at 5$\sigma$) for the objects at -40$^{\circ}$ < $\delta$ < +90$^{\circ}$. 

In addition, we also conducted dedicated follow-up radio observations using the Australian Telescope Compact Array (ATCA) for a sample of 18 UGSs (see Tab. A.1 in the appendices \href{https://zenodo.org/records/14543340}{see Zenodo}). 
These 18 sources were selected either because the survey images showed a potential, but not highly significant radio counterpart [a 2$\sigma$ $\leq$ SNR $\leq$ 4$\sigma$ detection from the analysis of the RACS and LOFAR (\citet[LOw-Frequency ARray Two-metre Sky Survey][]{LoTSS_2022}) images], due to a lack of radio coverage of their sky position, or because the survey images showed intriguing extended radio structures.
The data reduction and the main results of this campaign are described in the Appendix A (\href{https://zenodo.org/records/14543340}{see Zenodo}).

\section{Results}

\label{sec:results}

Among the 714 UGSs with Swift/XRT observation, we find that 274 objects have at least one X-ray detection with a SNR $\geq$3 inside the \textit{Fermi} error box.
For 193 of them, only one possible counterpart in the Fermi error box was found (hereafter UGS1; see an example of a UGS1 source in the top panel of Fig. \ref{fig:Xskymap_es}; the entire UGS1 sub-sample is shown in the appendices, \href{https://zenodo.org/records/14543340}{see Zenodo}.), while 81 have more than one potential X-ray counterparts (hereafter UGS2; see an example of a UGS2 source in the bottom panel of Fig. \ref{fig:Xskymap_es}). Details about the \textit{Swift}/XRT observations and all X-ray detections related to UGS1 and UGS2 are in Tab. \ref{tab:cont_X_1} and \ref{tab:cont_X_multi}, respectively. 
For each X-ray detection, we performed X-ray spectral data analysis (see Sec. \ref{sec:discussion}) and
in Tab. \ref{tab:spectra_analysis_1} and \ref{tab:spectra_analysis_multi}, we report the results of the XRT spectral fitting for UGS1 and UGS2. The spectral analysis allows us to estimate the source absorbed flux and to give a first, rough indication on the spectral slope.

We find that, over a total of 431 possible X-ray counterparts, 384 sources can be well fit with a simple absorbed power-law (0.8 $\leq$ $\chi^{2}_{\nu}$ $\leq$ 1.2). For the other 36 sources, the fitting statistics are very poor (less than 3 spectral points) and hence we only roughly estimated the absorbed flux by fixing the nH and the powerlaw index of the spectral model. For the further 11 X-ray sources, the spectral fits are unreliable ($\chi^{2}_{\nu}$ $\gg$ 1.2 or $\chi^{2}_{\nu}$ $\ll$ 0.8). 

\begin{figure}
\hspace{-0.5cm}
\centering
   \includegraphics[width=10.0truecm]{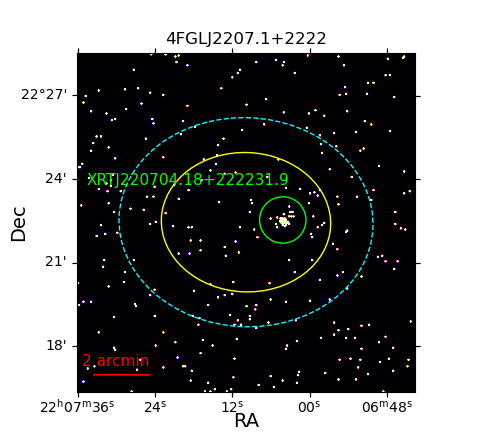}
    \includegraphics[width=10.0truecm]{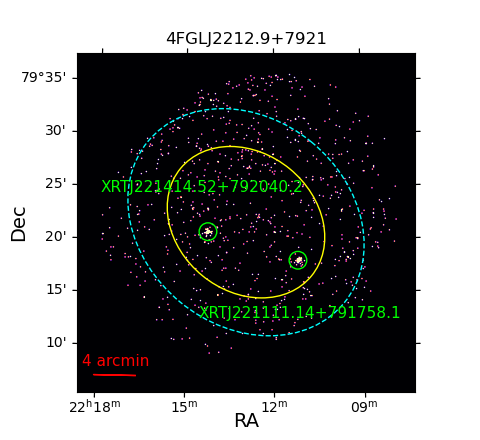}
\caption{\textit{Upper panel}: The X-ray skymap of 4FGL J22017.1+2222. The yellow and cyan ellipses indicate the $2\sigma$ and $3\sigma$ \textit{Fermi} $\gamma$-ray error regions, respectively. The X-ray detection is shown with a green circular region.\\
\textit{Bottom panel}: A X-ray skymap of 4FGL J2212.9+7921 with a colour legend as above.} 
\label{fig:Xskymap_es}
\end{figure}

\begin{table*}
\begin{center}
\caption{Summary of a subsample (shown as example) of \textit{Swift}/XRT detections within the 3$\sigma$ \textit{Fermi} error box of a list of UGS1 (UGS with a single potential X-ray counterpart), where a portion of the table is shown to demonstrate its form and content. A machine-readable version of the full table is available at the CDS.}
\resizebox{19cm }{!}{
\begin{tabular}{lrrcrrrr}
\hline 
4FGL Name (1) & $\gamma$-ray detection (2) & Swift exposure (3) & \textit{Swift}/XRT source (4) & RA (5) & DEC (6) & Positional error (7) & X-ray detection (8) \\   
          & ($\sigma$) & (ks)  & & (J2000) & (J2000) & (arcsec) & ($\sigma$)  \\  
\hline
4FGL J0003.6+3059 & 13.1 & 4.9 & XRT J000402.65+310219.8 & 1.0111 & 31.0388 & 4.6 & 3.7\\
4FGL J0004.4$-$4001 & 11.0 & 9.4 & XRT J000434.22$-$400034.73 & 1.1426 & -40.0096 & 2.9 & 7.7\\
4FGL J0006.6+4618 & 6.0 & 4.5 & XRT J000652.09+461813.9 & 1.7170 & 46.3039 & 6.4 & 3.5\\
4FGL J0009.1$-$5012 & 9.7 & 4.8 & XRT J000908.61$-$501000.9 & 2.2859 & -50.1669 & 6.2 & 3.9\\
4FGL J0022.0$-$5921 & 12.4 & 4.6 & XRT J002127.45$-$591946.3 & 5.3644 & -59.3295 & 6.0 & 4.1\\
4FGL J0023.2+8412 & 3.4 & 1.8 & XRT J002403.36+841352.6 & 6.0140 & 84.2313 & 8.0 & 3.4\\
4FGL J0023.6$-$4209 & 4.4 & 3.4 & XRT J002303.59$-$420509.6  & 5.7650 & -42.0860 & 2.9 & 5.0\\
4FGL J0025.4$-$4838 & 7.4 & 4.4 & XRT J002536.94$-$483810.9 & 6.4039 & -48.6364 & 3.0 & 5.6\\
4FGL J0026.1$-$0732 & 9.7 & 6.6 & XRT J002611.55$-$073116.0 & 6.5481 & -7.5211 & 2.4 & 22.6\\
4FGL J0027.0$-$1134 & 4.4 & 1.9 & XRT J002710.11$-$113638.7 & 6.7921 & -11.6108 & 3.2 & 5.1\\
\hline
\end{tabular}
}
\label{tab:cont_X_1}
\end{center}
\raggedright
\footnotesize{\textbf{Note.} 1) 4FGL Name; 2) $\gamma$-ray detection significance as reported in the 4FGL catalog;  3) \textit{Swift}/XRT exposure time; 4) Name of the \textit{Swift}/XRT source; 5-6) Coordinates of the X-ray source; 7) X-ray positional error radius; 8) Detection significance of the X-ray source.}
\end{table*}

\begin{table*}
\begin{center}
\caption{Summary of a subsample (shown as example) of \textit{Swift}/XRT detections within the 3$\sigma$ \textit{Fermi} error box of a list of UGS2 (UGS with more than one potential X-ray counterparts), where a portion of the table is shown to demonstrate its form and content. A machine-readable version of the full table is available at the CDS.}
\resizebox{19cm }{!}{
\begin{tabular}{lrrcrrrr}
\hline 
4FGL Name (1) & $\gamma$-ray detection (2) & Swift exposure (3) & \textit{Swift}/XRT source (4) & RA (5) & DEC (6) & Positional error (7) & X-ray detection (8) \\   
          & ($\sigma$) & (ks)  & & (J2000) & (J2000) & (arcsec) & ($\sigma$)  \\  
\hline
 &  &  & XRT J001708.63$-$460607.7 & 4.2859 & -46.1021 & 2.9 & 6.5\\
4FGL J0017.1$-$4605 & 7.2 & 9.8 & XRT J001750.81$-$460437.5  & 4.4617 & -46.0771 & 4.2 & 4.7\\
 &  &  & XRT J001705.00$-$460109.2 & 4.2709 & -46.0192 & 7.7 & 3.0\\
\hline
\multirow{2}{*}{4FGL J0031.0$-$2327} & \multirow{2}{*}{10.3} & \multirow{2}{*}{10.7} & XRT J003120.53$-$233400.7  & 7.8355 & -23.5669 & 2.5 & 19.5\\
 &  &  & XRT J003039.79$-$232821.2 & 7.6658 & -23.4725 & 6.6 & 3.6\\
\hline
 &  &  & XRT J004016.42$-$271912.3 & 10.0684 & -27.3201 & 2.0 & 25.2\\
 &  &  & XRT J004023.77$-$272254.2 & 10.0991 & -27.3817 & 7.0 & 5.6\\
4FGL J0040.2$-$2725 & 6.6 & 33.3 & XRT J004026.07$-$272116.1 & 10.1086 & -27.3545 & 7.0 & 5.6\\
 &  &  & XRT J004035.80$-$272240.7 & 10.1492 & -27.3780 & 7.6 & 4.4\\
 &  &  & XRT J003954.35$-$272516.1 & 9.9764 & -27.4211 & 6.0 & 3.9\\
\hline
 &  &  & XRT J004602.94$-$132422.2 & 11.5123 & -13.4062 & 7.0 & 7.7\\
 &  &  & XRT J004608.32$-$132213.6 & 11.5347 & -13.3704 & 3.5 & 5.2\\
\multirow{2}{*}{4FGL J0045.8$-$1324} & \multirow{2}{*}{5.9} & \multirow{2}{*}{5.7} & XRT J004611.48$-$132519.3 & 11.5479 & -13.4220 & 3.5 & 4.4\\
 &  &  & XRT J004539.41$-$132507.7 & 11.4142 & -13.4188 & 4.3 & 4.0\\
 &  &  & XRT J004602.97$-$131959.0 & 11.5124 & -13.3331 & 6.0 & 3.2\\
 &  &  & XRT J004555.20$-$132312.5 & 11.4800 & -13.3868 & 3.5 & 3.2\\
\hline
 & & & XRT J010226.89+095939.9 & 15.6121 & 9.9944 & 4.4 & 4.6\\
\multirow{2}{*}{4FGL J0102.3+1000} & \multirow{2}{*}{12.4} & \multirow{2}{*}{12.2} & XRT J010235.93+095832.4 & 15.6497 & 9.9757 & 6.0 & 3.7\\
 &  &  & XRT J010220.73+095848.9 & 15.5864 & 9.9803 & 6.0 & 3.4\\
 &  &  & XRT J010214.03+100258.4 & 15.5585 & 10.0496 & 3.6 & 3.3\\
\hline
\end{tabular}
}
\label{tab:cont_X_multi}
\end{center}
\raggedright
\footnotesize{\textbf{Note.} Labels are the same as in table \ref{tab:cont_X_1}.}
\end{table*}

\begin{figure}
\hspace{-0.5cm}
\centering
   \includegraphics[width=10truecm]{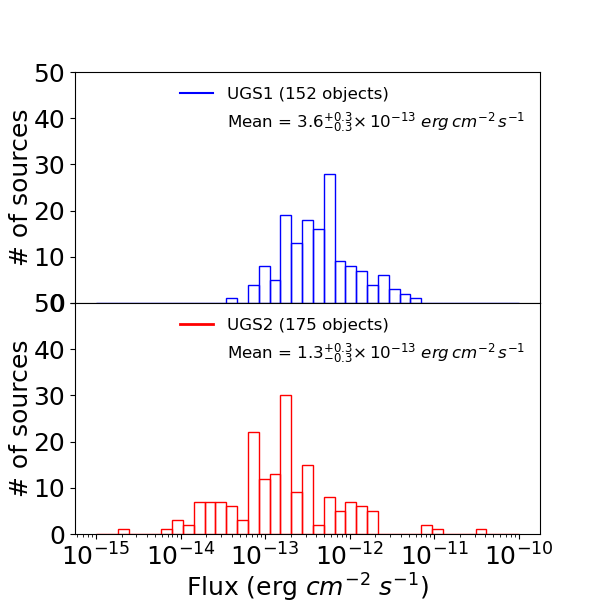}
\caption{Absorbed 0.3--10 keV flux distributions for UGS1 and UGS2. It is worth noting that sources with proper motion are excluded from the histograms.}
\label{fig:hist_flux}
\end{figure}

Comparing the distribution of absorbed 0.3--10 keV fluxes for UGS1 and UGS2 (see Fig. \ref{fig:hist_flux}, excluding sources that exhibit proper motion according to {\it GAIA}) and are therefore likely to be Galactic objects, we find that the UGS1 flux distribution follows a log-normal behaviour but the UGS2 distribution is less predictable, probably due to the dominance of spurious sources. The $\chi^2_{\nu}$ from fitting the X-ray fluxes of UGS2 with a log-normal function indicated that the model does not fit the data well. To avoid underestimating the uncertainties, we introduced a systematic error into the data for the UGS2 and repeated the fit.
The best-fitting results are reported in table \ref{tab:X-ray_flux}. 

Using the position of each X-ray detection, we the list the potential optical and radio counterparts for UGS1 and UGS2 in Tab. \ref{tab:MWL_count_1} and \ref{tab:MWL_count_multi}, providing the optical coordinates, the magnitude in the \textit{g} and \textit{r} bands, the optical and the radio flux. 
Finally we report the radio-loudness parameter \textit{R} defined as the ratio between the radio density flux, in the 2-4 GHz range, and the optical g-band density flux. Historically, sources with the \textit{R}$>$10 are defined as radio-loud sources, otherwise are radio-quiet \citep{Kellermann_1989}.

Over a total of 431 X-ray counterparts (193 for UGS1 and 238 for UGS2), we find that the \textit{GAIA} survey reports a significant ($> 3\sigma$) proper motion for 41 and 63 optical counterparts of UGS1 and UGS2, respectively, strongly supporting for a possible Galactic origin.

\subsection{UGS1 counterparts}

\begin{figure}
\hspace{-0.5cm}
\centering
   \includegraphics[width=10.0truecm]{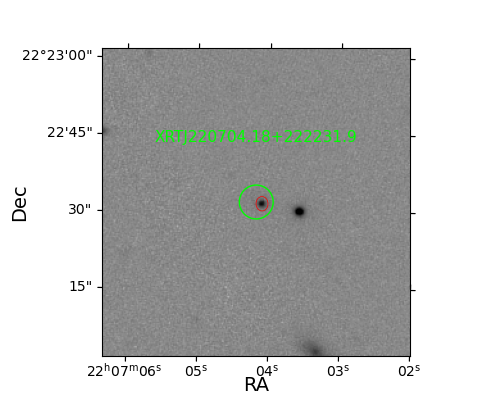}
   \includegraphics[width=10.0truecm]{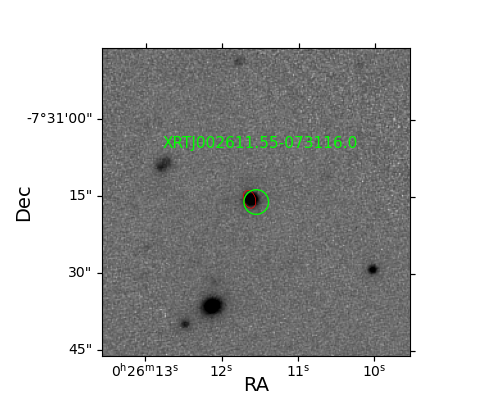}
\caption{\textit{Upper panel}: Optical r-band PanSTARRs image of 4FGL J2207.1+2222 counterpart. The green circle represent the error box of the X-ray counterpart and the red ellipses the error box of radio counterparts found within the VLASS catalog. \textit{Bottom panel}: Optical r-band PanSTARRs image of 4FGL J0026.1-0732 counterpart. Colour codes are as above.} 
\label{fig:Oskymap_2207}
\end{figure}

\begin{figure}
\hspace{-0.5cm}
\centering
   \includegraphics[width=10.0truecm]{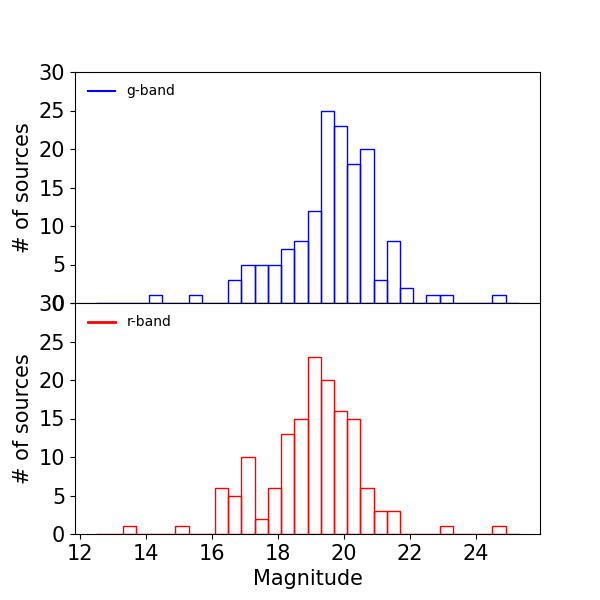}
\caption{Distribution of magnitudes in the g-band (top) and r-band (bottom) for UGS1. Note that sources with proper motion have been excluded from the histogram.} 
\label{fig:g_mag_diagram}
\end{figure}

All of 193 UGS1 X-ray counterparts are coincident with optical sources (see an example in Fig. \ref{fig:Oskymap_2207} and the entire sample in Fig. C.2, \href{https://zenodo.org/records/14543340}{(see Zenodo)}), that have \textit{g} magnitudes spanning from 6.3 to 24.9 (from 14.9 to 24.9 excluding sources with proper motion), while their \textit{r} magnitudes span from 6.7 to 24.9 (from 13.4 to 24.7 excluding sources with proper motion). The distribution of magnitudes in g-band and r-band are shown in  Figure \ref{fig:g_mag_diagram}.

Furthermore, 113 X-ray counterparts of the 193 UGS1 are also coincident with a radio source (105 excluding sources with proper motion).
33 UGS1 optical counterparts already have optical spectra available from the literature, which were presented in \citet[][]{Ulgiati_2024}.

All X-ray sources coincident with a radio source from the VLASS and RACS catalogs, and for which we have an estimate of the optical magnitude in the g-band, are radio-loud, except for 4FGL J0641.4+3349/PAN J064111.22+334459.7 with $R$=2. This source is classified as a low redshift QSO on the basis of the absolute magnitude and the emission lines detected in the optical spectrum \citep{monroe_2016, Ulgiati_2024}.

All the sources observed in the radio band using ATCA are radio-quiet except for five objects: 4FGL J0126.3$-$6746/SSS J012622.15$-$674623.0 ($R$ = 1601) and 4FGL J1709.4$-$2127/PAN J170936.70$-$212838.9 ($R$ = 90)\footnote{Both are extended radio sources also listed in the RACS and VLASS catalogs, respectively.}, 4FGL J0536.1$-$1205/PAN J053626.80$-$120652.1 (already listed in the VLASS catalog, $R$ = 216), 4FGL J0755.9-0515/PAN J075614.43$-$051718.9 ($R$ =16) and 4FGL J1415.9-1504/PAN J141546.17$-$150229.0 ($R$ = 13).

\begin{figure}
\hspace{-0.5cm}
\centering
   \includegraphics[width=10.0truecm]{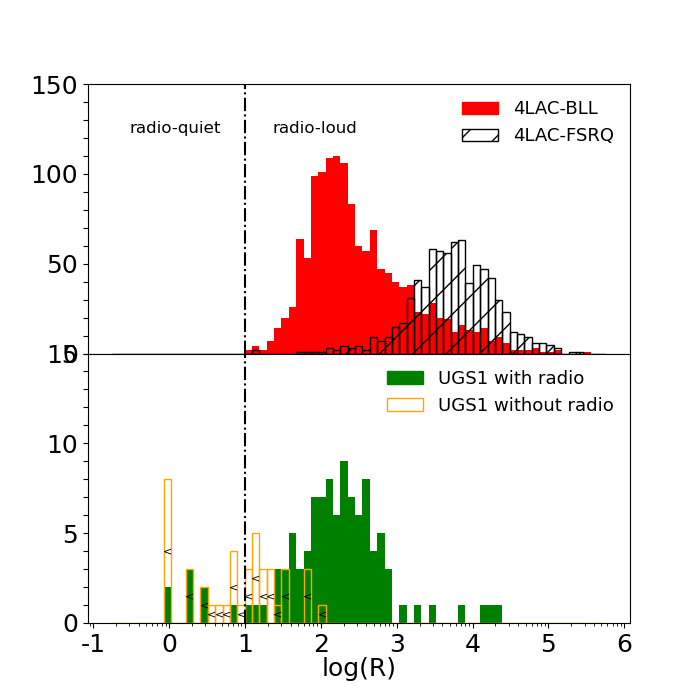}
\caption{\textit{Upper panel}: Distribution of the \textit{radio-loudness} parameter (log(R)) value for 1409 objects classified as BLL and 771 objects as FSRQ of the 4LAC catalog. \textit{Bottom panel}: Distribution of the \textit{radio-loudness} value for the counterparts of UGS1. It is worth to note that the empty orange bars represent the upper limits on R for UGS1 withot a radio counterpart. The black dashed vertical line represent the \textit{radio-loudness} parameter value (R = 10) that separates the radio-quiet from the radio-loud sources. It is worth noting that sources with proper motion are excluded from the histograms.} 
\label{fig:R_value}
\end{figure}

The other UGS1 X-ray sources that do not coincide with a radio source can be considered radio-quiet although for 30 of them the radio and \textit{R} upper limit are poorly constrained (R > 10). For a comparison between the distribution of the R value in the UGS1 sub-sample and in \textit{Fermi} blazars, see Fig. \ref{fig:R_value}.

\subsection{UGS2 counterparts}

Of the UGS2 sample, 54 of them could be associated with two potential X-ray counterparts, of which 7 have both counterparts exhibiting proper motion, and 18 have only one counterpart with proper motion. For the latter 18 UGSs, only 8 of them have at least one non-moving object coincident with a radio source. Additionally, 11 UGS2 objects have three potential X-ray counterparts, and 16 have more than three X-ray counterparts. Similar to our approach with UGS1 sources, all potential X-ray counterparts of UGS2 that coincide with a radio source cataloged in VLASS or RACS, and for which we have an estimate of the optical magnitude in the g-band, are found to be radio-loud, except 3: 4FGL J0159.0+3313/XRT J015905.35+331257.8, 4FGL J1008.2$-$1000/XRT J100848.62$-$095450.2, and 4FGL J1407.7$-$3017/XRT J140806.82$-$302353.7, which are radio quiet. Investigating the optical images and the literature on these objects, we found that XRT J015905.35+331257.8 is located very close to a star ($\sim$ 25 arcsec), which probably compromises the measurement of the optical magnitude; XRT J100848.62$-$095450.2 is a Seyfert 1 \citep[][]{Veron_2006} with a redshift of z=0.057253 \citep[][]{Jones_2009}; XRT J140806.82$-$302353.7 is a Seyfert 1 \citep[][]{Malizia_2012} with redshift z\,=\,0.023456 \citep[][]{Jones_2009}. The other UGS2 objects, which do not coincide with a radio object, are either radio-quiet or have an upper limit for R that exceeds the threshold of 10.

\begin{table*}
\begin{center}
\caption{Results of the X-ray spectral fitting for a list of UGS1 shown as example. A machine-readable version of the full table is available in the CDS.} 
\resizebox{16cm }{!}{
\begin{tabular}{lcccrrrr}
\hline 
4FGL Name (1) & \textit{Swift}/XRT source (2) & Count rate (3) & nH (4) & $\Gamma$  (5) & Norm. (6) & absorbed flux (7) & $\chi^{2}_{\nu}$(dof) (8)\\   
          &  & [$\times$10$^{-3}$] & [$\times$10$^{20}$] & & $10^{-5}$ & [$\times$10$^{-13}$] &\\  
\hline
4FGL J0003.6+3059 & XRT J000402.65+310219.8 & 4.64$_{-1.01}^{+1.01}$ & 6.8 & 2 & 4.3$_{-1.1}^{+1.1}$ & 2.0$_{-0.5}^{+0.5}$ & 0.01(1)\\
4FGL J0004.4$-$4001 & XRT J000434.22$-$400034.7 & 9.68$_{-1.04}^{+1.04}$ & 1.2 & 2.5$_{-0.2}^{+0.2}$ & 7.2$_{-0.9}^{+0.9}$ & 3.2$_{-0.4}^{+0.4}$ & 0.9(8)\\
4FGL J0006.6+4618 & XRT J000652.09+461813.9 & 4.19$_{-1.00}^{+1.00}$ & 11.0 & 2 & 5.8$_{-1.6}^{+1.6}$ & 2.5$_{-0.7}^{+0.7}$ & -\\
4FGL J0009.1$-$5012 & XRT J000908.61$-$501000.9 & 4.53$_{-1.01}^{+1.01}$ & 1.4 & 2 & 4.7$_{-1.5}^{+1.5}$ & 2.5$_{-0.8}^{+0.8}$ & -\\
4FGL J0022.0$-$5921 & XRT J002127.45$-$591946.3 & 6.14$_{-1.19}^{+1.19}$ & 1.4 & 2 & 6.1$_{-1.3}^{+1.3}$ & 3.2$_{-0.7}^{+0.7}$ & 0.7(2)\\
4FGL J0023.2+8412 & XRT J002403.36+841352.6 & 13.42$_{-2.84}^{+2.84}$ & 11.0 & 2 & 11.0$_{-2.9}^{+2.9}$ & 5.0$_{-1.4}^{+1.4}$ & 1.2(1)\\
4FGL J0023.6$-$4209 & XRT J002303.59-420509.6 & 12.60$_{-1.96}^{+1.96}$ & 1.4 & 1.2$_{-0.3}^{+0.3}$ & 11.0$_{-2.0}^{+2.0}$ & 12.6$_{-3.5}^{+4.4}$ & 0.8(3)\\
4FGL J0025.4$-$4838 & XRT J002536.94$-$483810.9 & 12.97$_{-1.77}^{+1.77}$ & 1.8 & 2.5$_{-0.2}^{+0.2}$ & 8.6$_{-1.4}^{+1.4}$ & 4.0$_{-0.7}^{+0.7}$ & 1.8(4)\\
4FGL J0026.1$-$0732 & XRT J002611.55$-$073116.0 & 94.75$_{-3.81}^{+3.81}$ & 4.0 & 2.2$_{-0.1}^{+0.1}$ & 77.0$_{-3.3}^{+3.3}$ & 31.6$_{-1.4}^{+1.5}$ & 0.7(66)\\
4FGL J0027.0$-$1134 & XRT J002710.11$-$113638.7 & 22.40$_{-3.50}^{+3.50}$ & 2.9 & 2.0$_{-0.3}^{+0.3}$ & 21.0$_{-3.6}^{+3.6}$ & 10.0$_{-2.1}^{+2.0}$ & 1.0(3)\\
\hline
\end{tabular}
}
\label{tab:spectra_analysis_1}
\end{center}
\raggedright
\footnotesize{\textbf{Note.} 1) 4FGL Name; 2) Name of the \textit{Swift}/XRT source; 3) \textit{Swift}/XRT count rate; 4) Equivalent hydrogen column density (cm$^{-2}$), provided by the \citet[][]{HI4PI_Coll_2016} database; 5) Power-law index; 6) Normalization factor (photons keV$^{-1}$ cm$^{-2}$ s$^{-1}$) at 1 keV; 7) Absorbed flux in the range 0.3--10 keV (erg cm$^{-2}$ s$^{-1}$); 8) $\chi^{2}_{\nu}$ (and the relative degrees of freedom).\\
The sources for which the $\chi^{2}_{\nu}$ does not appear in the table have a spectral data quality too poor to allow for spectral fitting. For these sources, we just convert the observed source count rate into a flux (see text).
}
\end{table*}

\begin{table*}
\begin{center}
\caption{Results of the X-ray spectral fitting for a list of UGS2 shown as example. A machine-readable version of the full table is available in the CDS.} 
\resizebox{16cm }{!}{
\begin{tabular}{lcccrrrr}
\hline 
4FGL Name (1) & \textit{Swift}/XRT source (2) & Count rate (3) & nH (4) & $\Gamma$  (5) & Norm. (6) & absorbed flux (7) & $\chi^{2}_{\nu}$(dof) (8)\\    
          &  & [$\times$10$^{-3}$] & [$\times$10$^{20}$] & & $10^{-5}$ & [$\times$10$^{-13}$] &\\   
\hline
 & XRT J001708.63$-$460607.7 & 6.44$_{-0.84}^{+0.84}$ & 1.6 & 1.8$_{-0.3}^{+0.3}$ & 5.0$_{-0.7}^{+0.7}$ & 3.2$_{-0.7}^{+0.7}$ & 0.6(6)\\
4FGL J0017.1$-$4605 & XRT J001750.81$-$460437.5 & 3.95$_{-0.66}^{+0.66}$ & 1.6 & -0.1$_{-0.6}^{+0.5}$ & 1.3$_{-0.5}^{+0.5}$ & 12.6$_{-6.1}^{+13.1}$ & 1.3(2)\\
 & XRT J001705.00$-$460109.2 & 1.47$_{-0.42}^{+0.42}$ & 1.6 & 2 & 1.1$_{-0.4}^{+0.4}$ &0.6$_{-0.2}^{+0.2}$ & -\\
\hline 
\multirow{2}{*}{4FGL J0031.0$-$2327} & XRT J003120.53$-$233400.7 & 44.80$_{-2.08}^{+2.08}$ & 1.7 & 1.8$_{-0.1}^{+0.1}$ & 34.0$_{-1.8}^{+1.8}$ & 20.0$_{-1.3}^{+1.4}$ & 0.9(50)\\
 & XRT J003039.79-232821.2 & 1.93$_{-0.48}^{+0.48}$ & 1.7 & 2 & 1.6$_{-0.5}^{+0.5}$ &0.8$_{-0.2}^{+0.2}$ & -\\
\hline 
 & XRT J004016.42$-$271912.3 & 22.59$_{-0.83}^{+0.83}$ & 1.4 & 2.7$_{-0.1}^{+0.1}$ & 16.0$_{-0.6}^{+0.6}$ & 6.3$_{-0.3}^{+0.3}$ & 1.1(72)\\
 & XRT J004023.77$-$272254.2 & 1.32$_{-0.22}^{+0.22}$ & 1.4 & 7.2$_{-0.3}^{+0.3}$ & 0.4$_{-0.1}^{+0.1}$ & 3.2$_{-0.9}^{+1.0}$ & 1.0(4)\\
\multirow{2}{*}{4FGL J0040.2$-$2725} & XRT J004026.07$-$272116.1 & 1.51$_{-0.23}^{+0.23}$ & 1.4 & 2.5$_{-0.2}^{+0.2}$ & 1.3$_{-0.2}^{+0.2}$ & 0.5$_{-0.1}^{+0.1}$ & 0.8(5)\\
 & XRT J004035.80$-$272240.7 & 0.91$_{-0.19}^{+0.19}$ & 1.4 & 1.1$_{-0.6}^{+0.6}$ & 0.5$_{-0.2}^{+0.1}$ & 0.6$_{-0.3}^{+0.5}$ & 2.4(2)\\
 & XRT J003954.35$-$272516.1 & 0.63$_{-0.16}^{+0.16}$ & 1.5 & 2 & 0.4$_{-0.1}^{+0.1}$ & 0.3$_{-0.1}^{+0.1}$ & 3.0(2)\\
\hline 
 & XRT J004602.94$-$132422.2 & 4.00$_{-0.44}^{+0.44}$ & 1.7 & 2.2$_{-0.2}^{+0.2}$ & 3.3$_{-0.4}^{+0.4}$ & 1.6$_{-0.2}^{+0.2}$ & 0.8(10)\\
 & XRT J004608.32$-$132213.6 & 1.90$_{-0.32}^{+0.32}$ & 1.7 & 1.1$_{-0.3}^{+0.3}$ & 1.1$_{-0.3}^{+0.3}$ & 1.6$_{-0.4}^{+0.5}$ & 1.1(4)\\
\multirow{2}{*}{4FGL J0045.8$-$1324} & XRT J004611.48$-$132519.3 & 1.68$_{-0.30}^{+0.30}$ & 1.7 & 1.7$_{-0.4}^{+0.4}$ & 1.8$_{-0.3}^{+0.3}$ & 1.3$_{-0.4}^{+0.8}$ & 1.2(3)\\
 & XRT J004539.41$-$132507.7 & 0.87$_{-0.23}^{+0.23}$ & 1.7 & 2 & 0.8$_{-0.2}^{+0.2}$ & 0.4$_{-0.1}^{+0.1}$ & 0.9(2)\\
 & XRT J004602.97$-$131959.0 & 0.39$_{-0.20}^{+0.20}$ & 1.8 & 2 & 0.5$_{-0.2}^{+0.2}$ &0.3$_{-0.1}^{+0.1}$ & -\\
 & XRT J004555.20$-$132312.5 & 0.73$_{-0.22}^{+0.22}$ & 1.7 & 2 & 0.6$_{-0.2}^{+0.2}$ & 0.3$_{-0.1}^{+0.1}$ & 2.4(1)\\
\hline 
 & XRT J010226.89+095939.9 & 0.47$_{-0.14}^{+0.14}$ & 4.3 & 2.4$_{-0.8}^{+0.7}$ & 0.6$_{-0.1}^{+0.1}$ & 0.3$_{-0.1}^{+0.2}$ & 1.1(2)\\
\multirow{2}{*}{4FGL J0102.3+1000} & XRT J010235.93+095832.4 & 0.26$_{-0.11}^{+0.11}$ & 4.3 & 2 & 0.3$_{-0.1}^{+0.1}$ &0.1$_{-0.1}^{+0.1}$ & -\\
 & XRT J010220.73+095848.9 & 0.36$_{-0.12}^{+0.12}$ & 4.3 & 2 & 0.4$_{-0.1}^{+0.1}$ & 0.2$_{-0.1}^{+0.1}$ & 0.5(2)\\
 & XRT J010214.03+100258.4 & 0.39$_{-0.12}^{+0.12}$ & 4.3 & 2 & 0.3$_{-0.1}^{+0.1}$ & 0.2$_{-0.1}^{+0.1}$ & 3.3(1)\\
\hline 
\end{tabular}
}
\label{tab:spectra_analysis_multi}
\end{center}
\raggedright
\footnotesize{\textbf{Note.} Labels are the same as in table \ref{tab:spectra_analysis_1}.}
\end{table*}

\begin{table*}
\begin{center}
\caption{MWL counterparts of a list of UGS1 shown as example. \\
A portion of the table is shown here to demonstrate its form and content. A machine-readable version of the full table is available at the CDS.} 
\resizebox{19cm }{!}{
\begin{tabular}{llllcccrrr}
\hline 
4FGL Name (1) &  \textit{Swift}/XRT source (2)        & Radio source (3)        & Optical source (4)      &  RA (5)     & DEC (6)  & gmag(rmag) (7) & $f^{\rm opt}_{\nu}$ (8) & $f^{\rm radio}_{\nu}$ (9) & $R$ (10)\\   
          & &  &   &  J2000  &  J2000  &  & [$\times$ 10$^{-28}$] & &  \\  
\hline
4FGL J0003.6+3059 & XRT J000402.65+310219.8 & - & PANJ000402.51+310222.0 & 1.0105 & 31.0394 & 21.2(20.1) & 1.2 & <0.3 & <29\\
4FGL J0004.4$-$4001 & XRT J000434.22$-$400034.7 & VLASS1QLCIRJ000434.19-400033.6 & DESJ000434.22$-$400035.1 & 1.1426 & -40.0098 & 16.9(16.2) & 63.1 & 32.5 & 51\\
4FGL J0006.6+4618 & XRT J000652.09+461813.9 & VLASS1QLCIRJ000652.31+461816.8 & PANJ000652.33+461817.0 & 1.7180 & 46.3047 & 19.9(19.6) & 4.0 & 2.6 & 64\\
4FGL J0009.1$-$5012 & XRT J000908.61$-$501000.9* & - & DESJ000908.64$-$501000.6 & 2.2860 & -50.1669 & 11.1(11.5) & 13182.6 & <0.6 & <1\\
4FGL J0022.0$-$5921 & XRT J002127.45$-$591946.3 & RACSJ002127.4$-$591949 & DESJ002127.51$-$591948.0 & 5.3646 & -59.3300 & 19.6(19.2) & 5.2 & 93.4 & 1780\\
4FGL J0023.2+8412 & XRT J002403.36+841352.6 & - & PANJ002405.61+841352.2 & 6.0234 & 84.2312 & 16.7(16.2) & 75.9 & <0.3 & <1\\
4FGL J0023.6$-$4209 & XRT J002303.59-420509.6 & - & DESJ002303.75$-$420508.6 & 5.7657 & -42.0857 & 15.6(15.0) & 208.9 & <0.3 & 1\\
4FGL J0025.4$-$4838 & XRT J002536.94$-$483810.9 & RACSJ002536.8$-$483810 & DESJ002536.92$-$483809.5 & 6.4039 & -48.6360 & 19.8(19.4) & 4.4 & 5.0 & 114\\
4FGL J0026.1$-$0732 & XRT J002611.55$-$073116.0 & VLASS1QLCIRJ002611.63$-$073115.5 & PANJ002611.64$-$073115.8 & 6.5485 & -7.5211 & 19.1(19.1) & 8.3 & 6.4 & 77\\
4FGL J0027.0$-$1134 & XRT J002710.11$-$113638.7 & VLASS1QLCIRJ002710.12$-$113637.5 & PANJ002710.12$-$113637.9 & 6.7922 & -11.6105 & 17.1(16.3) & 52.5 & 19.9 & 38\\
\hline
\end{tabular}
}
\label{tab:MWL_count_1}
\end{center}
\raggedright
\footnotesize{\textbf{Note.} 1) 4FGL Name; 2) X-ray source; 3) Radio source; 4) Optical source; 5) - 6) Coordinates of the optical counterpart. 7) g(r) magnitude of the optical source. 8) Optical density flux in unit of erg cm$^{-2}$ s$^{-1}$ Hz$^{-1}$. 9) Radio density flux (mJy); 10) \textit{Radio-loudness} parameter.\\
\textbf{*} Sources with proper motion.\\
 } 
\end{table*}

\begin{table*}
\begin{center}
\caption{MWL counterparts of a list of UGS2 shown as example. \\
A portion of the table is shown here to demonstrate its form and content. A machine-readable version of the full table is available at the CDS.}
\resizebox{19cm }{!}{
\begin{tabular}{llllcccrrr}
\hline 
4FGL Name (1) & \textit{Swift}/XRT source (2)        & Radio source (3)        & Optical source (4)      &  RA (5)     & DEC (6)  & gmag(rmag) (7) & $f^{\rm opt}_{\nu}$ (8) & $f^{\rm radio}_{\nu}$ (9) & $R$ (10)\\     
          & &  &   &  J2000  &  J2000  &  & [$\times$ 10$^{-28}$] & &  \\ 
\hline
 & XRT J001708.63$-$460607.7 & ATCAJ001708.65$-$460607.0 & DESJ001708.67$-$460606.9 & 4.2862 & -46.1019 & 18.3(18.1) & 17.4 & 0.1 & 1\\
4FGL J0017.1$-$4605 & XRT J001750.81$-$460437.5 & ATCAJ001750.67$-$460438.6 & DESJ001750.66$-$460438.5 & 4.4611 & -46.0774 & 17.8(17.9) & 27.5 & 0.5 & 3\\
 & XRT J001705.00$-$460109.2 & - & DESJ001704.83$-$460110.1 & 4.2702 & -46.0195 & 20.8(20.7) & 1.7 & <0.6 & <37\\
\hline
\multirow{2}{*}{4FGL J0031.0$-$2327} & XRT J003120.53$-$233400.7 & VLASS1QLCIRJ003120.56$-$233401.6 & PANJ003120.55$-$233401.2 & 7.8356 & -23.5670 & 20.1(19.0) & 3.3 & 3.9 & 118\\
 & XRT J003039.79-232821.2 & - & PANJ003039.76$-$232822.3 & 7.6656 & -23.4729 & 19.9(19.5) & 4.0 & <0.3 & <9\\
\hline
 & XRT J004016.42$-$271912.3 & VLASS1QLCIRJ004016.40$-$271911.5 & PANJ004016.41$-$271911.7 & 10.0684 & -27.3199 & 18.8(18.6) & 11.0 & 86.0 & 784\\
 & XRT J004023.77$-$272254.2 & - & PANJ004023.90$-$272255.4 & 10.0996 & -27.3821 & 22.6(20.8) & - & <0.3 & -\\
4FGL J0040.2$-$2725 & XRT J004026.07$-$272116.1* & - & DESJ004026.16$-$272117.6 & 10.1090 & -27.3549 & 14.8(14.2) & 436.5 & <0.3 & <1\\
 & XRT J004035.80$-$272240.7 & - & PANJ004035.87$-$272242.0 & 10.1494 & -27.3783 & 21.9(21.6) & 0.6 & <0.3 & <55\\
 & XRT J003954.35$-$272516.1 & - & PANJ003954.09-272513.5 & 9.9754 & -27.4204 & 21.2(21.1) & 1.2 & <0.3 & <29\\
\hline
 & XRT J004602.94$-$132422.2 & VLASS1QLCIRJ004602.82$-$132422.1 & PANJ004602.82$-$132422.4 & 11.5118 & -13.4062 & 18.7(18.7) & 12.0 & 8.1 & 68\\
 & XRT J004608.32$-$132213.6 & - & PANJ004608.44$-$132215.2 & 11.5352 & -13.3709 & 18.5(18.6) & 14.5 & <0.3 & <2\\
\multirow{2}{*}{4FGL J0045.8$-$1324} & XRT J004611.48$-$132519.3* & - & PANJ004611.47$-$132522.6 & 11.5478 & -13.4229 & 17.6(16.4) & 33.1 & <0.3 & <1\\
 & XRT J004539.41$-$132507.7 & VLASS1QLCIRJ004539.47$-$132508.0 & - & - & - & -(-) & - & 1.7 & -\\
 & XRT J004602.97$-$131959.0 & - & PANJ004602.68$-$131958.9 & 11.5112 & -13.3330 & 21.1(21.3) & 1.3 & <0.3 & <26\\
 & XRT J004555.20$-$132312.5 & - & PANJ004555.27$-$132314.3 & 11.4803 & -13.3873 & --(21.2) & - & <0.3 & -\\
\hline
 & XRT J010226.89+095939.9 & VLASS1QLCIRJ010226.69+095939.6 & - & - & - & -(-) & - & 6.5 & -\\
\multirow{2}{*}{4FGL J0102.3+1000} & XRT J010235.93+095832.4 & - & - & - & - & -(-) & - & <0.3 & -\\
 & XRT J010220.73+095848.9 & - & PANJ010220.59+095847.5 & 15.5858 & 9.9799 & 20.1(20.0) & 3.3 & <0.3 & <10\\
 & XRT J010214.03+100258.4 & - & - & - & - & -(-) & - & <0.3 & -\\
\hline
\end{tabular}
}
\label{tab:MWL_count_multi}
\end{center}
\raggedright
\footnotesize{\textbf{Note.} Labels are the same as in table \ref{tab:MWL_count_1}

 } 
\end{table*}

\begin{figure}
\hspace{-0.5cm}
\centering
   \includegraphics[width=10truecm]{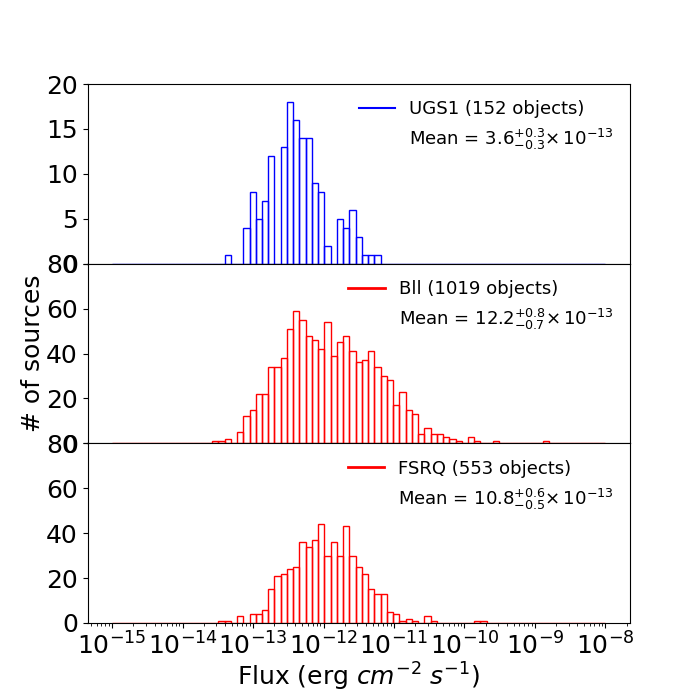}
\caption{Absorbed 0.3--10 keV flux distributions for UGS1 (top), 4FGL-DR4 BLL (centre) and 4FGL-DR4 FSRQ (bottom). It is worth noting that sources with proper motion are excluded from the histograms.}
\label{fig:hist_flux_1_blazar}
\end{figure}

\begin{table}
\begin{center}
\caption{Results of fits on X-ray flux distributions} 
\begin{tabular}{lcc}
\hline 
Sample (1) & Mean ($\mu$) (2) & Standard (3) \\
 & $\times$ 10$^{-13}$ & deviation($\sigma$)\\
 & erg cm$^{-2}$ s$^{-1}$ & $\times$ 10$^{-13}$\\
\hline
UGS1 & 3.6$^{+0.3}_{-0.3}$ & 3.2\\
UGS2 & 1.3$^{+0.3}_{-0.3}$ & 1.8\\
4FGL-DR4 BLL & 12.2$^{+0.8}_{-0.7}$ & 18.6\\
4FGL-DR4 FSRQ & 10.8$^{+0.6}_{-0.5}$ & 11.9\\
\hline
\end{tabular}
\label{tab:X-ray_flux}
\end{center}
\footnotesize{\textbf{Note.} Results of lognormal fits on the distributions of X-ray fluxes in the different samples, where $\mu$ is the mean. 1) Sample under investigation; 2) Mean energy flux in the range 0.3--10 keV (erg cm$^{-2}$ s$^{-1}$); 3) Standard deviation $\sigma$.}
\end{table}

\section{Discussion} 
\label{sec:discussion}

We compare the main flux properties of the UGS X-ray counterparts with the main class of extragalactic $\gamma$-ray emitters detected by \textit{Fermi} to characterize the nature of UGS counterparts and to test the goodness of our proposal. 
In particular, we confront the X-ray 0.3-10 keV flux distribution of UGS1 with those of the 4FGL-DR4 blazars (see Fig. \ref{fig:hist_flux_1_blazar}), where we carefully excluded Blazar Candidates of Uncertain type (BCU), since they could hide objects that are misclassified as blazars. As seen before, we adopted a log-normal function to describe the flux properties. However, we found that the flux distribution of 4FGL-DR4 BLL cannot be described log-normal distribution ($\chi^2_{\nu}$ $\gg$ 1 and p-value $\ll$ 0.05)\footnote{To address this, we introduce a systematic error into the data before performing the log-normal fit, to avoid underestimating the uncertainty of the model parameters.} The results of the fit are reported in table \ref{tab:X-ray_flux}. 

We note that the X-ray fluxes of UGS1 are generally lower than those of the 4FGL-DR4 blazars (expected values of the distributions in table \ref{tab:X-ray_flux}). A simple explanation for this behaviour is linked to the fact that UGSs are intrinsically weak $\gamma$-ray emitters, revealed by the ever-increasing exposure time (and hence the statistics) of the Fermi survey, which allows the detection of the sources with weakest fluxes. 
 
These probably belong to the blazar population at large distances and/or may be AGN with a low nucleus-to-host flux ratio.  

\begin{figure}
\hspace{-0.5cm}
\centering
   \includegraphics[width=9.0truecm]{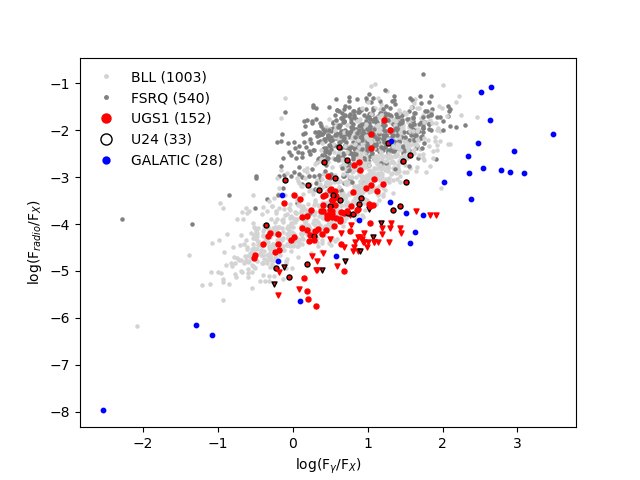}\\
   \includegraphics[width=9.0truecm]{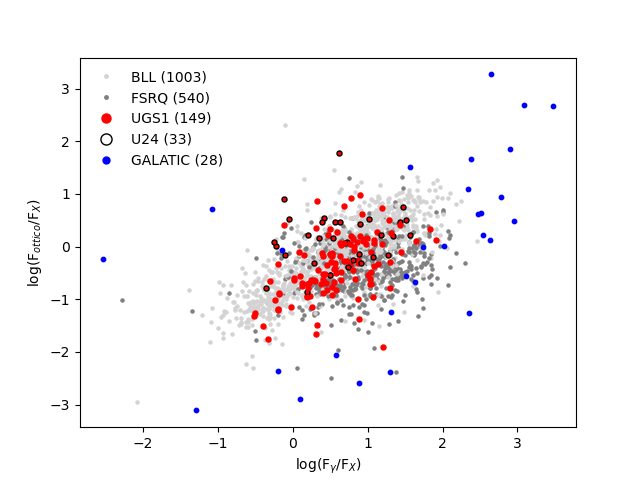}
\caption{log(F$_{\text{radio}}$/F$_{\text{X}}$) vs log(F$_{\gamma}$/F$_{\text{X}}$) (top) and log(F$_{\text{optical}}$/F$_{\text{X}}$) vs log(F$_{\gamma}$/F$_{\text{X}}$) (bottom) diagrams and  for 4FGL-DR4 BLL (light grey), 4FGL-DR4 FSRQ (grey), UGS1 (red) and 4FGL-DR4 Galactic sources (blue). Sources already analysed in \citep[][]{Ulgiati_2024} are underlined with a black edge around the points. It is worth noting that sources with proper motion or lacking g-band magnitude estimates are not included in the plot.}
\label{fig:diagn_UFO_blazar}
\end{figure}

\begin{figure}
\hspace{-0.5cm}
\centering
      \includegraphics[width=9truecm]{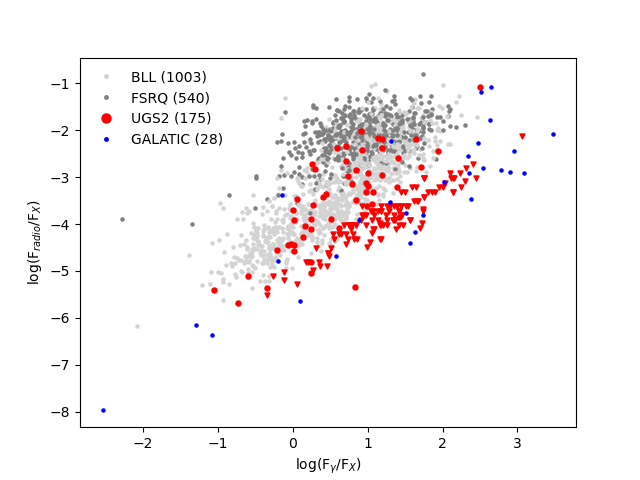}
      \includegraphics[width=9.0truecm]{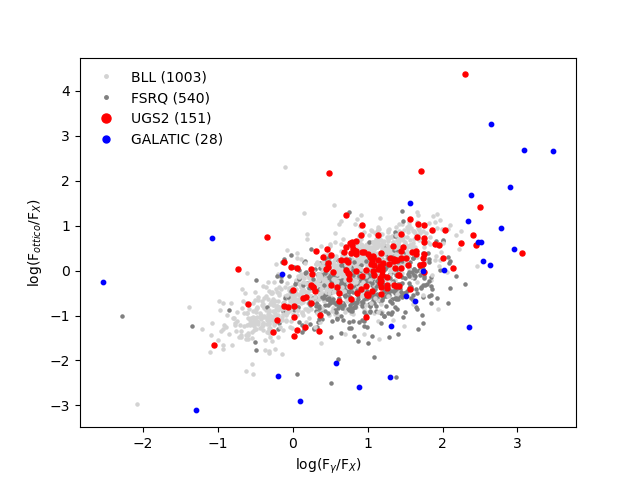}
\caption{log(F$_{\text{radio}}$/F$_{\text{X}}$) vs log(F$_{\gamma}$/F$_{\text{X}}$) (top) and log(F$_{\text{optical}}$/F$_{\text{X}}$) vs log(F$_{\gamma}$/F$_{\text{X}}$) (bottom) diagrams and  for 4FGL-DR4 BLL (light grey), 4FGL-DR4 FSRQ (grey), UGS2 (red) and 4FGL-DR4 Galactic sources (blue). It is worth noting that sources with proper motion, those lacking an optical counterpart, or those without g-band magnitude estimates are not included in the plot.}
\label{fig:diagn_UFO_deg_blazar}
\end{figure}

We further build the \textit{colour-colour diagrams} considering the ratios of the absorbed fluxes at different energy bands, in particular log(F$_{\text{radio}}$/F$_{\text{X}}$) and log(F$_{\text{optical}}$/F$_{\text{X}}$) as a function of log(F$_{\gamma}$/F$_{\text{X}}$) for UGS1 (see Fig. \ref{fig:diagn_UFO_blazar}) and UGS2 (see Fig. \ref{fig:diagn_UFO_deg_blazar}). In the same plots, we also included the 4FGL-DR4 blazars and 4FGL-DR4 Galactic sources (pulsars, pulsar wind nebulae, High Mass X-ray Binaries, Low Mass X-ray Binaries, and Supernova Remnants).
For the 4FGL-DR4 sources which are not covered by \textit{Swift}/XRT observations, we extract their X-ray fluxes from the 4XMM-DR13 catalog \citep[][]{Webb_2020}, in order to increase the number of sources with an estimated X-ray flux.

All Fermi AGNs ($\sim$ 98\% are blazars and hence radio-loud) occupy a compact region in the two \textit{colour-colour diagrams}, and it is possible to recognize two sub-regions: one populated by BLLs and one populated by FSRQs. 

The upper panel of Fig. \ref{fig:diagn_UFO_blazar} shows that a fraction of UGS1 (filled red points) occupies the same region of the 4FGL-DR4 blazars (grey points). These are the sources for which there is also a radio detection and are radio-loud. This supports the classification with the class of radio-loud AGN and, for 19 objects this is confirmed spectroscopically \citep{Ulgiati_2024}.

The remaining UGS1 without radio signal, and for which a radio flux upper limit is provided, are marked as filled red triangles. These sources, along with the UGS1 for which a potential radio counterpart was found using ATCA, considered as radio-quiet objects, are located in the lower edge of the locus occupied by the Fermi blazars. On the other hand, taking into account the second diagram displayed in the bottom panel of Fig. \ref{fig:diagn_UFO_blazar}, where the radio flux is not taken into account, these sources share the same region occupied by AGNs, suggesting a radio-quiet AGN nature. Of these sources, 10 of them were studied in \citet{Ulgiati_2024} and, indeed, they belong to the class of the Seyfert/QSO on the basis of their optical spectrum. 
The same analysis is performed for the UGS2 sample (see Fig. \ref{fig:diagn_UFO_deg_blazar}), where the degree of degeneracy of the number of the proposed counterparts leads to more speculative conclusions. 
Also in this case, we have the same trend found for the UGS1 for which the radio-loud sources occupy the same region of 4FGL-DR4 blazars while the radio-quiet ones are located in a distinct region.

\begin{figure}
\hspace{-0.5cm}
\centering
      \includegraphics[width=9.0truecm]{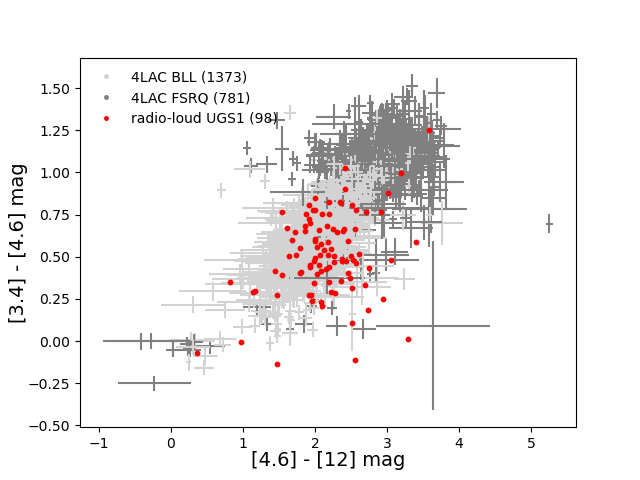}
      \includegraphics[width=9.0truecm]{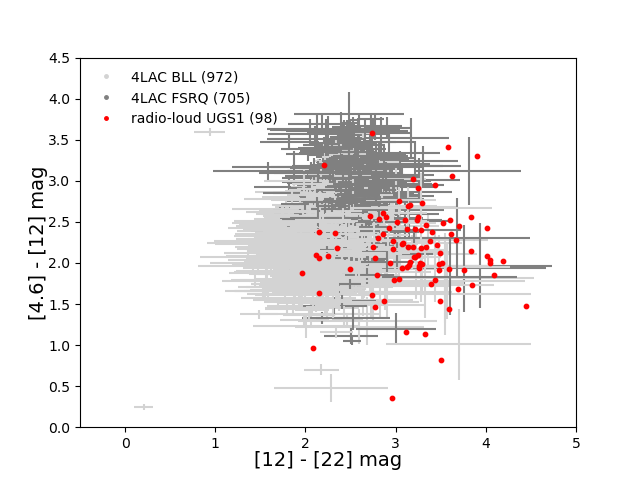}
      \includegraphics[width=9.0truecm]{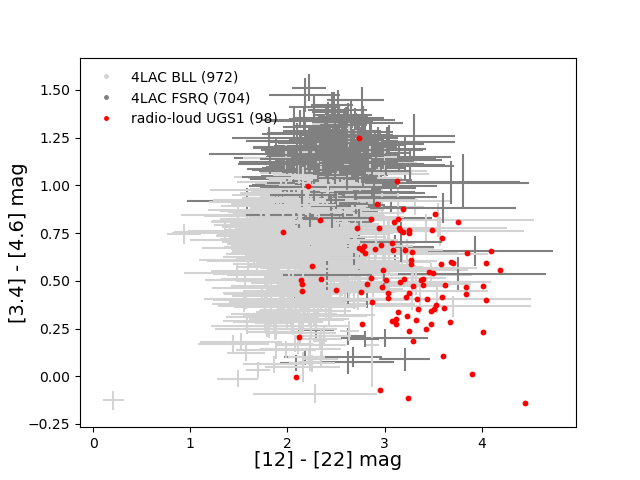}
\caption{[3.4]-[4.6]-[12] $\mu$m (top), [4.6]-[12]-[22] $\mu$m (centre) and [3.4]-[4.6]-[12]-[22] $\mu$m (bottom) IR colour diagrams for 4FGL-DR4 BLL (light grey), 4FGL-DR4 FSRQ (grey) and radio-loud UGS1 (red). It is worth noting that sources with proper motion are excluded from the histograms.}
\label{fig:color_IR_UGS1}
\end{figure}

\begin{figure}
\hspace{-0.5cm}
\centering
      \includegraphics[width=9.0truecm]{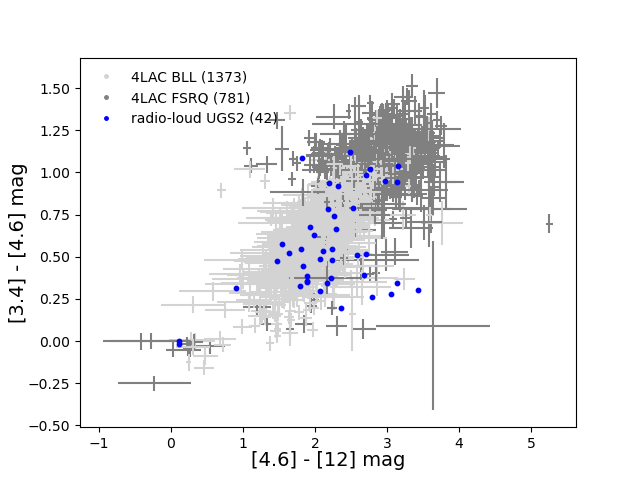}
      \includegraphics[width=9.0truecm]{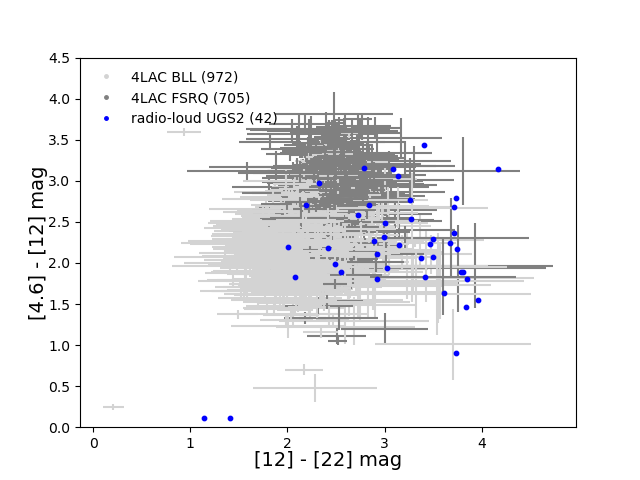}
      \includegraphics[width=9.0truecm]{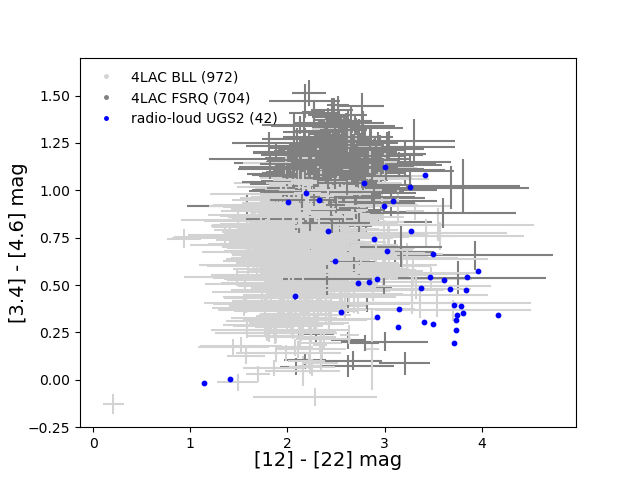}
\caption{[3.4]-[4.6]-[12] $\mu$m (top), [4.6]-[12]-[22] $\mu$m (centre) and [3.4]-[4.6]-[12]-[22] $\mu$m (bottom) IR colour diagrams for 4FGL-DR4 BLL (light grey), 4FGL-DR4 FSRQ (grey) and radio-loud UGS2 (blue). It is worth noting that sources with proper motion are excluded from the histograms.}
\label{fig:color_IR_UGS2}
\end{figure}

A detailed analysis of the radio-loud UGS, which are more likely to be blazars, was performed using the diagnostic diagram known as the WISE gamma-ray blazar strip \citep[][and references therein]{Massaro_2016}. This region in the IR colour-colour space is where gamma-ray-emitting blazars are typically located. The distributions of the radio-loud counterparts from UGS1 and UGS2 are shown in Fig. \ref{fig:color_IR_UGS1} and Fig. \ref{fig:color_IR_UGS2}, respectively.
The overlap of the radio-loud UGSs with the WISE blazar strip further supports the hypothesis that these sources are likely blazar candidates. However, definitive confirmation will require optical spectroscopy. We highlight that dedicated spectroscopic campaigns have partly been undertaken and are still ongoing, employing 10-meter-class telescopes (Paiano et al. in prep.). These efforts aim to obtain high-quality optical spectra to characterize their properties and confirm their nature. In addition, multi-wavelength investigations will focus on building the UGS SED to better understand their emission mechanisms. 
Such studies will offer insights into the physical processes powering these objects and will allow to distinguish between various classes of $\gamma$-ray emitters, representing an important step toward a more complete characterization of the UGS population.

\section{Conclusions}
\label{sec:conclusions}

The primary aim of this work is to create a catalog of possible counterparts and list their multi-wavelength emission fluxes. Detailed studies on individual sources are ongoing: optical spectroscopic campaigns are being conducted using optical telescopes of 10 meter class, as the Gran Telescopio Canarias (GTC), and multiwavelength study are focused on constructing the SEDs of these objects, with the aim of providing an overall view of their emission and gaining insights into the underlying physical processes \citep[][and Ulgiati et al in prep. and Paiano et al in prep.]{paiano2017_ufo1,paiano2019_ufo2,Ulgiati_2024}.

This paper is one piece of the global puzzle in the study of extra-galactic UGSs, a puzzle that will help shed light on these $\gamma$-ray emitters. New blazars and AGNs are discovered, further enriching our understanding of their population characteristics and how they fit into the larger framework of active galactic nuclei and their varied emission processes.

As a final remark, this paper presents a comprehensive list of potential lower-energy counterparts (X-ray, optical, and radio) to extra-galactic UGSs (|b| > 10$^{\circ}$) from the \textit{Fermi} 4FGL-DR4 catalog. The search for counterparts began in the X-ray band, analysing \textit{Swift}/XRT observations that covered the UGS regions. Of the 714 UGSs observed, 274 have at least one X-ray source within the 3$\sigma$ \textit{Fermi} error box, with a signal-to-noise ratio (SNR) $\geq$ 3. Among these, 193 have only one possible X-ray counterpart (denoted as UGS1), and 81 have multiple possible X-ray counterparts (denoted as UGS2).

All 193 X-ray counterparts of UGS1 are coincident with optical sources, of which 113 are also coincident with radio sources. All X-ray sources coincident with a radio source from the VLASS and RACS catalogs, and for which optical magnitudes in the g-band were available, are radio-loud, except for 4FGL J0641.4+3349/PAN J064111.22+334459.7, which has an R value of 2. The UGS1 sources not coincident with radio objects are likely radio-quiet sources. A subset of these sources was observed using the ATCA facility, and although several were detected, they are still confirmed to be radio-quiet.

Regarding the UGS2 sample, similarly to UGS1, all potential X-ray counterparts of UGS2s that coincide with a radio source cataloged in VLASS or RACS, and for which g-band optical magnitudes are available, are found to be radio-loud, except for 4FGL J0159.0+3313/XRT J015905.35+331257.8, 4FGL J1008.2$-$1000/XRT J100848.62$-$095450.2, and 4FGL J1407.7$-$3017/XRT J140806.82$-$302353.7. More studies on the broadband properties of these two sources are therefore needed to constrain their nature.

Finally, it was noted that UGSs exhibit lower flux across all bands compared to sources already associated and identified in the 4FGL-DR4 catalog. Using colour-colour diagrams, log(F$_{\text{radio}}$/F$_{\text{X}}$) and log(F$_{\text{optical}}$/F$_{\text{X}}$) as functions of log(F$_{\gamma}$/F$_{\text{X}}$), we observed that both UGS1 and UGS2 sources coincident with radio sources occupy the same regions of the diagram as blazars (or generally, radio-loud AGNs). The remaining UGS1 sources not coincident with radio objects occupy a distinct region populated by mainly radio-quiet AGNs \citep[see][]{Ulgiati_2024}. For UGS2 sources not coincident with radio objects, due to a high level of degeneracy and the likelihood of spurious sources, they occupy a less constrained region.

Furthermore, as observed in the IR colour-colour diagrams [3.4]-[4.6]-[12]-[22] $\mu$m, the radio-loud UGS objects overlap with the WISE blazar strip defined in \citet[][and references therein]{Massaro_2016}. This further supports the hypothesis that they are good candidates for being blazars. However, definitive confirmation will be provided by optical spectroscopy, which will help to better constrain their classification and confirm the blazar nature of the candidates.

\section*{Data availability}
The appendices for this article are available at Zenodo: \href{https://zenodo.org/records/14543340}{https://zenodo.org/records/14543340}.\\
Tables 1-2-3-4-5-6 are only available in electronic form at the CDS via anonymous ftp to \href{cdsarc.u-strasbg.fr}{cdsarc.u-strasbg.fr} (130.79.128.5) or via \href{http://cdsweb.u-strasbg.fr/cgi-bin/qcat?J/A+A/}{http://cdsweb.u-strasbg.fr/cgi-bin/qcat?J/A+A/}.

\begin{acknowledgements}
FP acknowledges support from INAF Grant OBIWAN.
CP acknowledges support from PRIN MUR SEAWIND (2022Y2T94C) funded by European Union - NextGenerationEU and INAF Grant BLOSSOM. MDS aknowledges support from INAF Grant ACEB-BANANA. This work has been partially supported by the ASI-INAF program I/004/11/6.

\end{acknowledgements}

\bibliographystyle{aa} 
\bibliography{biblio}

\begin{thebibliography}{96}
\expandafter\ifx\csname natexlab\endcsname\relax\def\natexlab#1{#1}\fi

\bibitem[{{Abbott} {et~al.}(2021){Abbott}, {Adam{\'o}w}, {Aguena}, {Allam},
  {Amon}, {Annis}, {Avila}, {Bacon}, {Banerji}, {Bechtol}, {Becker}, \& {Linea
  Science Server}}]{Abbott_2021}
{Abbott}, T.~M.~C., {Adam{\'o}w}, M., {Aguena}, M., {et~al.} 2021, \apjs, 255,
  20

\bibitem[{{Abdo} {et~al.}(2010){Abdo}, {Ackermann}, {Ajello}, {Atwood},
  {Baldini}, {Ballet}, {Barbiellini}, {Bastieri}, {Baughman}, {Bechtol}, \&
  {Fermi LAT Collaboration}}]{Abdo_2010}
{Abdo}, A.~A., {Ackermann}, M., {Ajello}, M., {et~al.} 2010, Science, 328, 725

\bibitem[{{Abdo} {et~al.}(2013){Abdo}, {Ajello}, {Allafort}, {Baldini},
  {Ballet}, {Barbiellini}, {Baring}, {Bastieri}, {Belfiore}, {Bellazzini}, \&
  {Bhattacharyya}}]{Abdo_2013}
{Abdo}, A.~A., {Ajello}, M., {Allafort}, A., {et~al.} 2013, \apjs, 208, 17

\bibitem[{{Abdollahi} {et~al.}(2020){Abdollahi}, {Acero}, {Ackermann},
  {Ajello}, {Atwood}, {Axelsson}, {Baldini}, {Ballet}, {Barbiellini}, \&
  {Bastieri}}]{4FGL_cat}
{Abdollahi}, S., {Acero}, F., {Ackermann}, M., {et~al.} 2020, \apjs, 247, 33

\bibitem[{{Abdollahi} {et~al.}(2022){Abdollahi}, {Acero}, {Baldini}, {Ballet},
  {Bastieri}, {Bellazzini}, {Berenji}, {Berretta}, \& {Bissaldi}}]{4FGL_DR3}
{Abdollahi}, S., {Acero}, F., {Baldini}, L., {et~al.} 2022, \apjs, 260, 53

\bibitem[{{Acciari} {et~al.}(2019){Acciari}, {Ansoldi}, {Antonelli}, {Arbet
  Engels}, {Baack}, {Babi{\'c}}, {}, {Banerjee}, {Barres de Almeida}, {Barrio},
  {Becerra Gonz{\'a}lez}, {Bednarek}, {Bellizzi}, {Bernardini}, {Berti},
  {Besenrieder}, {Bhattacharyya}, {Bigongiari}, {Biland}, {Blanch}, {Bonnoli},
  {Busetto}, {Carosi}, {Ceribella}, {Chai}, {Cikota}, {Colak}, {Colin},
  {Colombo}, {Contreras}, {Cortina}, {Covino}, {D'Elia}, {Da Vela}, {Dazzi},
  {De Angelis}, {De Lotto}, {Delfino}, {Delgado}, {Di Pierro}, {Do Souto
  Espi{\~n}eira}, {Dom{\'\i}nguez}, {Dominis Prester}, {Dorner}, {Doro},
  {Elsaesser}, {Fallah Ramazani}, {Fattorini}, {Fern{\'a}ndez-Barral},
  {Ferrara}, {Fidalgo}, {Foffano}, {Fonseca}, {Font}, {Fruck}, {Galindo},
  {Gallozzi}, {Garc{\'\i}a L{\'o}pez}, {Garczarczyk}, {Gasparyan}, {Gaug},
  {Godinovi{\'c}}, {}, {Green}, {Guberman}, {Hadasch}, {Hahn}, {Hassan},
  {Herrera}, {Hoang}, {Hrupec}, {Inoue}, {Ishio}, {Iwamura}, {Kubo}, {Kushida},
  {Lamastra}, {Lelas}, {Leone}, {Lindfors}, {Lombardi}, {Longo}, {L{\'o}pez},
  {L{\'o}pez-Coto}, {L{\'o}pez-Oramas}, {Machado de Oliveira Fraga}, {Maggio},
  {Majumdar}, {Makariev}, {Mallamaci}, {Maneva}, {Manganaro}, {Mannheim},
  {Maraschi}, {Mariotti}, {Mart{\'\i}nez}, {Masuda}, {Mazin}, {Mi{\'c}},
  {anovi{\'c}}, {}, {Miceli}, {Minev}, {Miranda}, {Mirzoyan}, {Molina},
  {Moralejo}, {Morcuende}, {Moreno}, {Moretti}, {Munar-Adrover}, {Neustroev},
  {Niedzwiecki}, {Nievas Rosillo}, {Nigro}, {Nilsson}, {Ninci}, {Nishijima},
  {Noda}, {Nogu{\'e}s}, {N{\"o}the}, {Paiano}, {Palacio}, {Palatiello},
  {Paneque}, {Paoletti}, {Paredes}, {Pe{\~n}il}, {Peresano}, {Persic}, {Prada
  Moroni}, {Prandini}, {Puljak}, {Rhode}, {Rib{\'o}}, {Rico}, {Righi},
  {Rugliancich}, {Saha}, {Sahakyan}, {Saito}, {Satalecka}, {Schweizer},
  {Sitarek}, {{\v{S}}nidari{\'c}}, {}, {Sobczynska}, {Somero}, {Stamerra},
  {Strom}, {Strzys}, {Suri{\'c}}, {}, {Tavecchio}, {Temnikov}, {Terzi{\'c}},
  {}, {Teshima}, {Torres-Alb{\`a}}, {Tsujimoto}, {van Scherpenberg}, {Vanzo},
  {V{\'a}zquez Acosta}, {Vovk}, {Will}, {Zari{\'c}}, \& {}}]{Acciari_2019}
{Acciari}, V.~A., {Ansoldi}, S., {Antonelli}, L.~A., {et~al.} 2019, \mnras,
  486, 4233

\bibitem[{{Acero} {et~al.}(2015){Acero}, {Ackermann}, {Ajello}, {Albert},
  {Atwood}, {Axelsson}, {Baldini}, {Ballet}, {Barbiellini}, {Bastieri}, \&
  {Fermi-LAT Collaboration}}]{Acero_2015}
{Acero}, F., {Ackermann}, M., {Ajello}, M., {et~al.} 2015, \apjs, 218, 23

\bibitem[{{Acero} {et~al.}(2013){Acero}, {Donato}, {Ojha}, {Stevens},
  {Edwards}, {Ferrara}, {Blanchard}, {Lovell}, \& {Thompson}}]{Acero_2013}
{Acero}, F., {Donato}, D., {Ojha}, R., {et~al.} 2013, \apj, 779, 133

\bibitem[{{Ackermann} {et~al.}(2011){Ackermann}, {Ajello}, {Allafort},
  {Angelakis}, {Axelsson}, {Baldini}, {Ballet}, {Barbiellini}, {Bastieri}, \&
  {Bellazzini}}]{Ackermann_2011}
{Ackermann}, M., {Ajello}, M., {Allafort}, A., {et~al.} 2011, \apj, 741, 30

\bibitem[{{Ackermann} {et~al.}(2012){Ackermann}, {Ajello}, {Allafort},
  {Antolini}, {Baldini}, {Ballet}, {Barbiellini}, {Bastieri}, {Bellazzini}, \&
  {Berenji}}]{Ackermann_2012}
{Ackermann}, M., {Ajello}, M., {Allafort}, A., {et~al.} 2012, \apj, 753, 83

\bibitem[{{Ahumada} {et~al.}(2020){Ahumada}, {Allende Prieto}, {Almeida},
  {Anders}, {Anderson}, {Andrews}, {Anguiano}, \& {Arcodia}}]{Ahumada_2020}
{Ahumada}, R., {Allende Prieto}, C., {Almeida}, A., {et~al.} 2020, \apjs, 249,
  3

\bibitem[{{Ajello} {et~al.}(2014){Ajello}, {Romani}, {Gasparrini}, {Shaw},
  {Bolmer}, {Cotter}, {Finke}, {Greiner}, {Healey}, {King}, {Max-Moerbeck},
  {Michelson}, {Potter}, {Rau}, {Readhead}, {Richards}, \&
  {Schady}}]{Ajello_2014}
{Ajello}, M., {Romani}, R.~W., {Gasparrini}, D., {et~al.} 2014, \apj, 780, 73

\bibitem[{{Angioni} {et~al.}(2017){Angioni}, {Grandi}, {Torresi}, {Vignali}, \&
  {Kn{\"o}dlseder}}]{Angioni_2017}
{Angioni}, R., {Grandi}, P., {Torresi}, E., {Vignali}, C., \& {Kn{\"o}dlseder},
  J. 2017, in American Institute of Physics Conference Series, Vol. 1792, 6th
  International Symposium on High Energy Gamma-Ray Astronomy, 050006

\bibitem[{{Arnaud}(2022)}]{Arnaud_2022}
{Arnaud}, K. 2022, in AAS/High Energy Astrophysics Division, Vol.~54, AAS/High
  Energy Astrophysics Division, 203.02

\bibitem[{{Arnaud}(1996)}]{Arnaud_1996}
{Arnaud}, K.~A. 1996, in Astronomical Society of the Pacific Conference Series,
  Vol. 101, Astronomical Data Analysis Software and Systems V, ed. G.~H.
  {Jacoby} \& J.~{Barnes}, 17

\bibitem[{{Atwood} {et~al.}(2009){Atwood}, {Abdo}, {Ackermann}, {Althouse},
  {Anderson}, {Axelsson}, {Baldini}, {Ballet}, {Band}, \&
  {Barbiellini}}]{Atwood2009}
{Atwood}, W.~B., {Abdo}, A.~A., {Ackermann}, M., {et~al.} 2009, \apj, 697, 1071

\bibitem[{{Ballet} {et~al.}(2023){Ballet}, {Bruel}, {Burnett}, {Lott}, \& {The
  Fermi-LAT collaboration}}]{4FGL_DR4}
{Ballet}, J., {Bruel}, P., {Burnett}, T.~H., {Lott}, B., \& {The Fermi-LAT
  collaboration}. 2023, arXiv e-prints, arXiv:2307.12546

\bibitem[{{Bruzewski} {et~al.}(2021){Bruzewski}, {Schinzel}, {Taylor}, \&
  {Petrov}}]{Bruzewski_2021}
{Bruzewski}, S., {Schinzel}, F.~K., {Taylor}, G.~B., \& {Petrov}, L. 2021,
  \apj, 914, 42

\bibitem[{{Bruzewski} {et~al.}(2022){Bruzewski}, {Schinzel}, {Taylor}, \&
  {Petrov}}]{Bruzewski_2022}
{Bruzewski}, S., {Schinzel}, F.~K., {Taylor}, G.~B., \& {Petrov}, L. 2022,
  VizieR Online Data Catalog, J/ApJ/914/42

\bibitem[{{Cerruti}(2020)}]{Cerruti2020}
{Cerruti}, M. 2020, in Journal of Physics Conference Series, Vol. 1468, Journal
  of Physics Conference Series, 012094

\bibitem[{{Cerruti} {et~al.}(2011){Cerruti}, {Zech}, {Boisson}, \&
  {Inoue}}]{Cerruti2011}
{Cerruti}, M., {Zech}, A., {Boisson}, C., \& {Inoue}, S. 2011, in SF2A-2011:
  Proceedings of the Annual meeting of the French Society of Astronomy and
  Astrophysics, ed. G.~{Alecian}, K.~{Belkacem}, R.~{Samadi}, \&
  D.~{Valls-Gabaud}, 555--558

\bibitem[{{Chambers} {et~al.}(2016){Chambers}, {Magnier}, {Metcalfe},
  {Flewelling}, {Huber}, {Waters}, {Denneau}, {Draper}, {Farrow}, \&
  {Finkbeiner}}]{Chambers_2016}
{Chambers}, K.~C., {Magnier}, E.~A., {Metcalfe}, N., {et~al.} 2016, arXiv
  e-prints, arXiv:1612.05560

\bibitem[{{Cheung} \& {Fermi LAT Collaboration}(2010)}]{Cheung_2010}
{Cheung}, C.~C. \& {Fermi LAT Collaboration}. 2010, in AAS/High Energy
  Astrophysics Division, Vol.~11, AAS/High Energy Astrophysics Division \#11,
  30.07

\bibitem[{{Costamante}(2020)}]{Costamante_2020}
{Costamante}, L. 2020, in Multifrequency Behaviour of High Energy Cosmic
  Sources - XIII. 3-8 June 2019. Palermo, 35

\bibitem[{{Costamante} {et~al.}(2018){Costamante}, {Cutini}, {Tosti},
  {Antolini}, \& {Tramacere}}]{costomante2018}
{Costamante}, L., {Cutini}, S., {Tosti}, G., {Antolini}, E., \& {Tramacere}, A.
  2018, \mnras, 477, 4749

\bibitem[{{D'Abrusco} {et~al.}(2013){D'Abrusco}, {Massaro}, {Paggi}, {Masetti},
  {Tosti}, {Giroletti}, \& {Smith}}]{Dabrusco_2013}
{D'Abrusco}, R., {Massaro}, F., {Paggi}, A., {et~al.} 2013, \apjs, 206, 12

\bibitem[{{di Mauro}(2018)}]{Dimauro_2018}
{di Mauro}, M. 2018, in Fourteenth Marcel Grossmann Meeting - MG14, ed.
  M.~{Bianchi}, R.~T. {Jansen}, \& R.~{Ruffini}, 3098--3104

\bibitem[{{Doert} \& {Errando}(2014)}]{Doert_2014}
{Doert}, M. \& {Errando}, M. 2014, \apj, 782, 41

\bibitem[{{Evans} {et~al.}(2009){Evans}, {Beardmore}, {Page}, {Osborne},
  {O'Brien}, {Willingale}, {Starling}, {Burrows}, \& {et al.,}}]{Evans_2009}
{Evans}, P.~A., {Beardmore}, A.~P., {Page}, K.~L., {et~al.} 2009, \mnras, 397,
  1177

\bibitem[{{Evans} {et~al.}(2020){Evans}, {Page}, {Osborne}, {Beardmore},
  {Willingale}, {Burrows}, {Kennea}, {Perri}, {Capalbi}, {Tagliaferri}, \&
  {Cenko}}]{Evans_2020}
{Evans}, P.~A., {Page}, K.~L., {Osborne}, J.~P., {et~al.} 2020, \apjs, 247, 54

\bibitem[{{Falcone} {et~al.}(2011){Falcone}, {Stroh}, {Ferrara}, {Grove}, {Saz
  Parkinson}, {Burrows}, {Cheung}, {Donato}, {Gehrels}, \&
  {Kennea}}]{Falcone_2011}
{Falcone}, A., {Stroh}, M., {Ferrara}, E., {et~al.} 2011, in AAS/High Energy
  Astrophysics Division, Vol.~12, AAS/High Energy Astrophysics Division \#12,
  4.03

\bibitem[{{Falcone} {et~al.}(2014){Falcone}, {Stroh}, \&
  {Pryal}}]{Falcone_2014}
{Falcone}, A., {Stroh}, M., \& {Pryal}, M. 2014, in American Astronomical
  Society Meeting Abstracts, Vol. 223, American Astronomical Society Meeting
  Abstracts \#223, 301.05

\bibitem[{{Falomo} {et~al.}(2014){Falomo}, {Pian}, \& {Treves}}]{falomo_2014}
{Falomo}, R., {Pian}, E., \& {Treves}, A. 2014, \aapr, 22, 73

\bibitem[{{Fronte} {et~al.}(2023){Fronte}, {Mazzon}, {Metruccio}, {Munaretto},
  {Doro}, {Giommi}, {Viale}, \& {Barres de Almeida}}]{Fronte_2023}
{Fronte}, L., {Mazzon}, B., {Metruccio}, F., {et~al.} 2023, in Journal of
  Physics Conference Series, Vol. 2429, Journal of Physics Conference Series,
  012045

\bibitem[{{Gao} {et~al.}(2019){Gao}, {Fedynitch}, {Winter}, \&
  {Pohl}}]{Gao2019}
{Gao}, S., {Fedynitch}, A., {Winter}, W., \& {Pohl}, M. 2019, Nature Astronomy,
  3, 88

\bibitem[{{Gehrels} {et~al.}(2004){Gehrels}, {Chincarini}, {Giommi}, {Mason},
  {Nousek}, {Wells}, {White}, {Barthelmy}, {Burrows}, {Cominsky}, {Hurley},
  {Marshall}, {M{\'e}sz{\'a}ros}, {Roming}, {Angelini}, {Barbier}, {Belloni},
  {Campana}, {Caraveo}, {Chester}, {Citterio}, {Cline}, {Cropper}, {Cummings},
  {Dean}, {Feigelson}, {Fenimore}, {Frail}, {Fruchter}, {Garmire}, {Gendreau},
  {Ghisellini}, {Greiner}, {Hill}, {Hunsberger}, {Krimm}, {Kulkarni}, {Kumar},
  {Lebrun}, {Lloyd-Ronning}, {Markwardt}, {Mattson}, {Mushotzky}, {Norris},
  {Osborne}, {Paczynski}, {Palmer}, {Park}, {Parsons}, {Paul}, {Rees},
  {Reynolds}, {Rhoads}, {Sasseen}, {Schaefer}, {Short}, {Smale}, {Smith},
  {Stella}, {Tagliaferri}, {Takahashi}, {Tashiro}, {Townsley}, {Tueller},
  {Turner}, {Vietri}, {Voges}, {Ward}, {Willingale}, {Zerbi}, \&
  {Zhang}}]{Gehrels_2004}
{Gehrels}, N., {Chincarini}, G., {Giommi}, P., {et~al.} 2004, \apj, 611, 1005

\bibitem[{{Ghisellini} {et~al.}(2017){Ghisellini}, {Righi}, {Costamante}, \&
  {Tavecchio}}]{Ghisellini_2017}
{Ghisellini}, G., {Righi}, C., {Costamante}, L., \& {Tavecchio}, F. 2017,
  \mnras, 469, 255

\bibitem[{{Giommi} {et~al.}(2013){Giommi}, {Padovani}, \&
  {Polenta}}]{Giommi_2013}
{Giommi}, P., {Padovani}, P., \& {Polenta}, G. 2013, \mnras, 431, 1914

\bibitem[{{Goad} {et~al.}(2007){Goad}, {Osborne}, {Beardmore}, \&
  {Evans}}]{Goad_2007}
{Goad}, M.~R., {Osborne}, J.~P., {Beardmore}, A.~P., \& {Evans}, P.~A. 2007,
  GRB Coordinates Network, 7133, 1

\bibitem[{{Grandi}(2012)}]{Grandi_2012}
{Grandi}, P. 2012, in International Journal of Modern Physics Conference
  Series, Vol.~8, International Journal of Modern Physics Conference Series,
  25--30

\bibitem[{{Hale} {et~al.}(2021){Hale}, {McConnell}, {Thomson}, {Lenc}, {Heald},
  {Hotan}, {Leung}, {Moss}, {Murphy}, {Pritchard}, {Sadler}, {Stewart}, \&
  {Whiting}}]{Hale_2021}
{Hale}, C.~L., {McConnell}, D., {Thomson}, A.~J.~M., {et~al.} 2021, \pasa, 38,
  e058

\bibitem[{{Hambly} {et~al.}(2004){Hambly}, {Read}, {Mann}, {Sutorius}, {Bond},
  {MacGillivray}, {Williams}, \& {Lawrence}}]{Hambly_2004}
{Hambly}, N., {Read}, M., {Mann}, R., {et~al.} 2004, in Astronomical Society of
  the Pacific Conference Series, Vol. 314, Astronomical Data Analysis Software
  and Systems (ADASS) XIII, ed. F.~{Ochsenbein}, M.~G. {Allen}, \& D.~{Egret},
  137

\bibitem[{{HI4PI Collaboration} {et~al.}(2016){HI4PI Collaboration}, {Ben
  Bekhti}, {Fl{\"o}er}, {Keller}, {Kerp}, {Lenz}, {Winkel}, {Bailin},
  {Calabretta}, {Dedes}, {Ford}, {Gibson}, {Haud}, {Janowiecki}, {Kalberla},
  {Lockman}, {McClure-Griffiths}, {Murphy}, {Nakanishi}, {Pisano}, \&
  {Staveley-Smith}}]{HI4PI_Coll_2016}
{HI4PI Collaboration}, {Ben Bekhti}, N., {Fl{\"o}er}, L., {et~al.} 2016, \aap,
  594, A116

\bibitem[{{J{\"a}rvel{\"a}} {et~al.}(2021){J{\"a}rvel{\"a}}, {Berton}, \&
  {Crepaldi}}]{Jarvela_2021}
{J{\"a}rvel{\"a}}, E., {Berton}, M., \& {Crepaldi}, L. 2021, Frontiers in
  Astronomy and Space Sciences, 8, 147

\bibitem[{{Jones} {et~al.}(2009){Jones}, {Read}, {Saunders}, {Colless},
  {Jarrett}, {Parker}, {Fairall}, \& {Mauch}}]{Jones_2009}
{Jones}, D.~H., {Read}, M.~A., {Saunders}, W., {et~al.} 2009, \mnras, 399, 683

\bibitem[{{Kaur} {et~al.}(2019a){Kaur}, {Falcone}, \& {Stroh}}]{Kaur_2019}
{Kaur}, A., {Falcone}, A.~D., \& {Stroh}, M. 2019a, in AAS/High Energy
  Astrophysics Division, Vol.~17, AAS/High Energy Astrophysics Division, 106.09

\bibitem[{{Kaur} {et~al.}(2019b){Kaur}, {Falcone}, {Stroh}, {Kennea}, \&
  {Ferrara}}]{Kaur_2019b}
{Kaur}, A., {Falcone}, A.~D., {Stroh}, M.~D., {Kennea}, J.~A., \& {Ferrara},
  E.~C. 2019b, \apj, 887, 18

\bibitem[{{Kaur} {et~al.}(2023){Kaur}, {Kerby}, \& {Falcone}}]{Kaur_2023}
{Kaur}, A., {Kerby}, S., \& {Falcone}, A.~D. 2023, \apj, 943, 167

\bibitem[{{Kellermann} {et~al.}(1989){Kellermann}, {Sramek}, {Schmidt},
  {Shaffer}, \& {Green}}]{Kellermann_1989}
{Kellermann}, K.~I., {Sramek}, R., {Schmidt}, M., {Shaffer}, D.~B., \& {Green},
  R. 1989, \aj, 98, 1195

\bibitem[{{Kerby} {et~al.}(2021){Kerby}, {Kaur}, {Falcone}, {Eskenasy},
  {Hancock}, {Stroh}, {Ferrara}, {Ray}, {Kennea}, \& {Grove}}]{Kerby_2021}
{Kerby}, S., {Kaur}, A., {Falcone}, A.~D., {et~al.} 2021, \apj, 923, 75

\bibitem[{{Lacy} {et~al.}(2020){Lacy}, {Baum}, {Chandler}, {Chatterjee},
  {Clarke}, {Deustua}, {English}, {Farnes}, {Gaensler}, {Gugliucci},
  {Hallinan}, {Kent}, \& et~al.}]{Lacy_2020}
{Lacy}, M., {Baum}, S.~A., {Chandler}, C.~J., {et~al.} 2020, \pasp, 132, 035001

\bibitem[{{Landi} {et~al.}(2015){Landi}, {Bassani}, {Stephen}, {Masetti},
  {Malizia}, \& {Ubertini}}]{Landi_2015}
{Landi}, R., {Bassani}, L., {Stephen}, J.~B., {et~al.} 2015, \aap, 581, A57

\bibitem[{{Li} {et~al.}(2018){Li}, {Hou}, {Strader}, {Takata}, {Kong},
  {Chomiuk}, {Swihart}, {Hui}, \& {Cheng}}]{Li_2018}
{Li}, K.-L., {Hou}, X., {Strader}, J., {et~al.} 2018, \apj, 863, 194

\bibitem[{{Malizia} {et~al.}(2012){Malizia}, {Bassani}, {Bazzano}, {Bird},
  {Masetti}, {Panessa}, {Stephen}, \& {Ubertini}}]{Malizia_2012}
{Malizia}, A., {Bassani}, L., {Bazzano}, A., {et~al.} 2012, \mnras, 426, 1750

\bibitem[{{Mao} \& {Yu}(2013)}]{Mao_2013}
{Mao}, Z. \& {Yu}, Y.-W. 2013, Research in Astronomy and Astrophysics, 13, 952

\bibitem[{{Marchesini} {et~al.}(2020){Marchesini}, {Paggi}, {Massaro},
  {Masetti}, {D'Abrusco}, \& {Andruchow}}]{Marchesini_2020}
{Marchesini}, E.~J., {Paggi}, A., {Massaro}, F., {et~al.} 2020, \aap, 638, A128

\bibitem[{{Massaro} {et~al.}(2016){Massaro}, {{\'A}lvarez Crespo}, {D'Abrusco},
  {Landoni}, {Masetti}, {Ricci}, {Milisavljevic}, {Paggi}, {Chavushyan},
  {Jim{\'e}nez-Bail{\'o}n}, {Pati{\~n}o-{\'A}lvarez}, {Strader}, {Chomiuk}, {La
  Franca}, {Smith}, \& {Tosti}}]{Massaro_2016}
{Massaro}, F., {{\'A}lvarez Crespo}, N., {D'Abrusco}, R., {et~al.} 2016, \apss,
  361, 337

\bibitem[{{Massaro} {et~al.}(2015){Massaro}, {Landoni}, {D'Abrusco},
  {Milisavljevic}, {Paggi}, {Masetti}, {Smith}, \& {Tosti}}]{Massaro_2015}
{Massaro}, F., {Landoni}, M., {D'Abrusco}, R., {et~al.} 2015, \aap, 575, A124

\bibitem[{{Mirabal} {et~al.}(2012){Mirabal}, {Fr{\'\i}as-Martinez}, {Hassan},
  \& {Fr{\'\i}as-Martinez}}]{Mirabal_2012}
{Mirabal}, N., {Fr{\'\i}as-Martinez}, V., {Hassan}, T., \&
  {Fr{\'\i}as-Martinez}, E. 2012, \mnras, 424, L64

\bibitem[{{Monroe} {et~al.}(2016){Monroe}, {Prochaska}, {Tejos}, {Worseck},
  {Hennawi}, {Schmidt}, {Tumlinson}, \& {Shen}}]{monroe_2016}
{Monroe}, T.~R., {Prochaska}, J.~X., {Tejos}, N., {et~al.} 2016, \aj, 152, 25

\bibitem[{{Nasa High Energy Astrophysics Science Archive Research
  Center}(2014)}]{Nasa_2014}
{Nasa High Energy Astrophysics Science Archive Research Center}. 2014,
  {HEAsoft: Unified Release of FTOOLS and XANADU}, Astrophysics Source Code
  Library, record ascl:1408.004

\bibitem[{{Nolan} {et~al.}(2012){Nolan}, {Abdo}, {Ackermann}, {Ajello},
  {Allafort}, {Antolini}, {Atwood}, {Axelsson}, {Baldini}, \&
  {Ballet}}]{Nolan_2012}
{Nolan}, P.~L., {Abdo}, A.~A., {Ackermann}, M., {et~al.} 2012, \apjs, 199, 31

\bibitem[{{Nori} {et~al.}(2014){Nori}, {Giroletti}, {Massaro}, {D'Abrusco},
  {Paggi}, {Tosti}, \& {Funk}}]{Nori_2014}
{Nori}, M., {Giroletti}, M., {Massaro}, F., {et~al.} 2014, \apjs, 212, 3

\bibitem[{{Padovani} {et~al.}(2017){Padovani}, {Alexander}, {Assef}, {De
  Marco}, {Giommi}, {Hickox}, {Richards}, {Smol{\v{c}}i{\'c}},
  {Hatziminaoglou}, {Mainieri}, \& {Salvato}}]{Padovani_2017}
{Padovani}, P., {Alexander}, D.~M., {Assef}, R.~J., {et~al.} 2017, \aapr, 25, 2

\bibitem[{{Paiano} {et~al.}(2017{\natexlab{a}}){Paiano}, {Falomo},
  {Franceschini}, {Treves}, \& {Scarpa}}]{paiano2017_ufo1}
{Paiano}, S., {Falomo}, R., {Franceschini}, A., {Treves}, A., \& {Scarpa}, R.
  2017{\natexlab{a}}, \apj, 851, 135

\bibitem[{{Paiano} {et~al.}(2019){Paiano}, {Falomo}, {Treves}, {Franceschini},
  \& {Scarpa}}]{paiano2019_ufo2}
{Paiano}, S., {Falomo}, R., {Treves}, A., {Franceschini}, A., \& {Scarpa}, R.
  2019, \apj, 871, 162

\bibitem[{{Paiano} {et~al.}(2021){Paiano}, {Falomo}, {Treves}, {Padovani},
  {Giommi}, \& {Scarpa}}]{Paiano_2021b}
{Paiano}, S., {Falomo}, R., {Treves}, A., {et~al.} 2021, \mnras, 504, 3338

\bibitem[{{Paiano} {et~al.}(2023){Paiano}, {Falomo}, {Treves}, {Padovani},
  {Giommi}, {Scarpa}, {Bisogni}, \& {Marini}}]{Paiano_2023}
{Paiano}, S., {Falomo}, R., {Treves}, A., {et~al.} 2023, \mnras, 521, 2270

\bibitem[{{Paiano} {et~al.}(2020){Paiano}, {Falomo}, {Treves}, \&
  {Scarpa}}]{Paiano_2020}
{Paiano}, S., {Falomo}, R., {Treves}, A., \& {Scarpa}, R. 2020, \mnras, 497, 94

\bibitem[{{Paiano} {et~al.}(2017{\natexlab{b}}){Paiano}, {Franceschini}, \&
  {Stamerra}}]{Paiano_SED}
{Paiano}, S., {Franceschini}, A., \& {Stamerra}, A. 2017{\natexlab{b}}, \mnras,
  468, 4902

\bibitem[{{Paiano} {et~al.}(2017{\natexlab{c}}){Paiano}, {Landoni}, {Falomo},
  {Treves}, \& {Scarpa}}]{Paiano_2017b}
{Paiano}, S., {Landoni}, M., {Falomo}, R., {Treves}, A., \& {Scarpa}, R.
  2017{\natexlab{c}}, \apj, 844, 120

\bibitem[{{Paiano} {et~al.}(2017{\natexlab{d}}){Paiano}, {Landoni}, {Falomo},
  {Treves}, {Scarpa}, \& {Righi}}]{Paiano_2017a}
{Paiano}, S., {Landoni}, M., {Falomo}, R., {et~al.} 2017{\natexlab{d}}, \apj,
  837, 144

\bibitem[{{Paliya} {et~al.}(2015){Paliya}, {Stalin}, \&
  {Ravikumar}}]{Paliya_2015}
{Paliya}, V.~S., {Stalin}, C.~S., \& {Ravikumar}, C.~D. 2015, \aj, 149, 41

\bibitem[{{Petrov} {et~al.}(2013){Petrov}, {Mahony}, {Edwards}, {Sadler},
  {Schinzel}, \& {McConnell}}]{Petrov_2013}
{Petrov}, L., {Mahony}, E.~K., {Edwards}, P.~G., {et~al.} 2013, \mnras, 432,
  1294

\bibitem[{{Rieger}(2017)}]{Rieger_2017}
{Rieger}, F.~M. 2017, in American Institute of Physics Conference Series, Vol.
  1792, 6th International Symposium on High Energy Gamma-Ray Astronomy, 020008

\bibitem[{{Rodrigues} {et~al.}(2019){Rodrigues}, {Gao}, {Fedynitch},
  {Palladino}, \& {Winter}}]{Rodrigues2019}
{Rodrigues}, X., {Gao}, S., {Fedynitch}, A., {Palladino}, A., \& {Winter}, W.
  2019, \apjl, 874, L29

\bibitem[{{Salvetti} {et~al.}(2017{\natexlab{a}}){Salvetti}, {Chiaro}, {La
  Mura}, \& {Thompson}}]{Salvetti_2017b}
{Salvetti}, D., {Chiaro}, G., {La Mura}, G., \& {Thompson}, D.~J.
  2017{\natexlab{a}}, \mnras, 470, 1291

\bibitem[{{Salvetti} {et~al.}(2017{\natexlab{b}}){Salvetti}, {Mignani}, {De
  Luca}, {Marelli}, {Pallanca}, {Breeveld}, {H{\"u}semann}, {Belfiore},
  {Becker}, \& {Greiner}}]{Salvetti_2017a}
{Salvetti}, D., {Mignani}, R.~P., {De Luca}, A., {et~al.} 2017{\natexlab{b}},
  \mnras, 470, 466

\bibitem[{{Schinzel} {et~al.}(2017){Schinzel}, {Petrov}, {Taylor}, \&
  {Edwards}}]{Schinzel_2017}
{Schinzel}, F.~K., {Petrov}, L., {Taylor}, G.~B., \& {Edwards}, P.~G. 2017,
  \apj, 838, 139

\bibitem[{{Schinzel} {et~al.}(2015){Schinzel}, {Petrov}, {Taylor}, {Mahony},
  {Edwards}, \& {Kovalev}}]{Schinzel_2015}
{Schinzel}, F.~K., {Petrov}, L., {Taylor}, G.~B., {et~al.} 2015, \apjs, 217, 4

\bibitem[{{Seabroke} {et~al.}(2021){Seabroke}, {Fabricius}, {Teyssier},
  {Sartoretti}, {Katz}, {Cropper}, {Antoja}, {Benson}, {Smith}, {Dolding},
  {Gosset}, {Panuzzo}, {Th{\'e}venin}, {Allende Prieto}, \&
  {Blomme}}]{Seabroke_2021}
{Seabroke}, G.~M., {Fabricius}, C., {Teyssier}, D., {et~al.} 2021, \aap, 653,
  A160

\bibitem[{{Shaw} {et~al.}(2012){Shaw}, {Romani}, {Cotter}, {Healey},
  {Michelson}, {Readhead}, {Richards}, {Max-Moerbeck}, {King}, \&
  {Potter}}]{Shaw_2012}
{Shaw}, M.~S., {Romani}, R.~W., {Cotter}, G., {et~al.} 2012, \apj, 748, 49

\bibitem[{{Shaw} {et~al.}(2013){Shaw}, {Romani}, {Cotter}, {Healey},
  {Michelson}, {Readhead}, {Richards}, {Max-Moerbeck}, {King}, \&
  {Potter}}]{Shaw_2013}
{Shaw}, M.~S., {Romani}, R.~W., {Cotter}, G., {et~al.} 2013, \apj, 764, 135

\bibitem[{{Shimwell} {et~al.}(2022){Shimwell}, {Hardcastle}, {Tasse}, {Best},
  {R{\"o}ttgering}, {Williams}, {Botteon}, {Drabent}, {Mechev}, {Shulevski},
  {van Weeren}, {Bester}, {Br{\"u}ggen}, {Brunetti}, {Callingham}, {Chy{\.z}y},
  {Conway}, {Dijkema}, {Duncan}, {de Gasperin}, {Hale}, {Haverkorn}, {Hugo},
  {Jackson}, {Mevius}, {Miley}, {Morabito}, {Morganti}, {Offringa}, {Oonk},
  {Rafferty}, {Sabater}, {Smith}, {Schwarz}, {Smirnov}, {O'Sullivan},
  {Vedantham}, {White}, {Albert}, {Alegre}, {Asabere}, {Bacon}, {Bonafede},
  {Bonnassieux}, {Brienza}, {Bilicki}, {Bonato}, {Calistro Rivera}, {Cassano},
  {Cochrane}, {Croston}, {Cuciti}, {Dallacasa}, {Danezi}, {Dettmar}, {Di
  Gennaro}, {Edler}, {En{\ss}lin}, {Emig}, {Franzen}, {Garc{\'\i}a-Vergara},
  {Grange}, {G{\"u}rkan}, {Hajduk}, {Heald}, {Heesen}, {Hoang}, {Hoeft},
  {Horellou}, {Iacobelli}, {Jamrozy}, {Jeli{\'c}}, {Kondapally}, {Kukreti},
  {Kunert-Bajraszewska}, {Magliocchetti}, {Mahatma}, {Ma{\l}ek}, {Mandal},
  {Massaro}, {Meyer-Zhao}, {Mingo}, {Mostert}, {Nair}, {Nakoneczny},
  {Nikiel-Wroczy{\'n}ski}, {Orr{\'u}}, {Pajdosz-{\'S}mierciak}, {Pasini},
  {Prandoni}, {van Piggelen}, {Rajpurohit}, {Retana-Montenegro}, {Riseley},
  {Rowlinson}, {Saxena}, {Schrijvers}, {Sweijen}, {Siewert}, {Timmerman},
  {Vaccari}, {Vink}, {West}, {Wo{\l}owska}, {Zhang}, \& {Zheng}}]{LoTSS_2022}
{Shimwell}, T.~W., {Hardcastle}, M.~J., {Tasse}, C., {et~al.} 2022, \aap, 659,
  A1

\bibitem[{{Stickel} {et~al.}(1991){Stickel}, {Padovani}, {Urry}, {Fried}, \&
  {Kuehr}}]{Stickel_1991}
{Stickel}, M., {Padovani}, P., {Urry}, C.~M., {Fried}, J.~W., \& {Kuehr}, H.
  1991, \apj, 374, 431

\bibitem[{{Stocke} {et~al.}(1991){Stocke}, {Morris}, {Gioia}, {Maccacaro},
  {Schild}, {Wolter}, {Fleming}, \& {Henry}}]{Stocke_1991}
{Stocke}, J.~T., {Morris}, S.~L., {Gioia}, I.~M., {et~al.} 1991, \apjs, 76, 813

\bibitem[{{Stroh} \& {Falcone}(2013)}]{Stroh_2013}
{Stroh}, M.~C. \& {Falcone}, A.~D. 2013, \apjs, 207, 28

\bibitem[{{Takahashi} {et~al.}(2012){Takahashi}, {Kataoka}, {Nakamori},
  {Maeda}, {Makiya}, {Totani}, {Cheung}, {Stawarz}, {Guillemot}, {Freire}, \&
  {Cognard}}]{Takahashi_2012}
{Takahashi}, Y., {Kataoka}, J., {Nakamori}, T., {et~al.} 2012, \apj, 747, 64

\bibitem[{{Ulgiati} {et~al.}(2024){Ulgiati}, {Paiano}, {Treves}, {Falomo},
  {Sbarufatti}, {Pintore}, {Russell}, \& {Cusumano}}]{Ulgiati_2024}
{Ulgiati}, A., {Paiano}, S., {Treves}, A., {et~al.} 2024, \mnras, 530, 4626

\bibitem[{{Urry} \& {Padovani}(1995)}]{Urry_1995}
{Urry}, C.~M. \& {Padovani}, P. 1995, \pasp, 107, 803

\bibitem[{{V{\'e}ron-Cetty} \& {V{\'e}ron}(2006)}]{Veron_2006}
{V{\'e}ron-Cetty}, M.~P. \& {V{\'e}ron}, P. 2006, \aap, 455, 773

\bibitem[{{Webb} {et~al.}(2020){Webb}, {Coriat}, {Traulsen}, {Ballet}, {Motch},
  {Carrera}, {Koliopanos}, {Authier}, {de la Calle}, {Ceballos}, {Colomo},
  {Chuard}, {Freyberg}, {Garcia}, {Kolehmainen}, {Lamer}, {Lin}, {Maggi},
  {Michel}, {Page}, {Page}, {Perea-Calderon}, {Pineau}, {Rodriguez}, {Rosen},
  {Santos Lleo}, {Saxton}, {Schwope}, {Tom{\'a}s}, {Watson}, \&
  {Zakardjian}}]{Webb_2020}
{Webb}, N.~A., {Coriat}, M., {Traulsen}, I., {et~al.} 2020, \aap, 641, A136

\bibitem[{{Wilms} {et~al.}(2000){Wilms}, {Allen}, \& {McCray}}]{Wilms_2000}
{Wilms}, J., {Allen}, A., \& {McCray}, R. 2000, \apj, 542, 914

\bibitem[{{Wright} {et~al.}(2010){Wright}, {Eisenhardt}, {Mainzer}, {Ressler},
  {Cutri}, {Jarrett}, {Kirkpatrick}, {Padgett}, {McMillan}, {Skrutskie},
  {Stanford}, {Cohen}, {Walker}, {Mather}, {Leisawitz}, {Gautier}, {McLean},
  {Benford}, {Lonsdale}, {Blain}, {Mendez}, {Irace}, {Duval}, {Liu}, {Royer},
  {Heinrichsen}, {Howard}, {Shannon}, {Kendall}, {Walsh}, {Larsen}, {Cardon},
  {Schick}, {Schwalm}, {Abid}, {Fabinsky}, {Naes}, \& {Tsai}}]{Wright_2010}
{Wright}, E.~L., {Eisenhardt}, P. R.~M., {Mainzer}, A.~K., {et~al.} 2010, \aj,
  140, 1868

\bibitem[{{Wu}(2018)}]{Wu_2018}
{Wu}, H. K.~J. 2018, PhD thesis, Rheinische Friedrich Wilhelms University of
  Bonn, Germany

\bibitem[{{Ye} {et~al.}(2023){Ye}, {Zeng}, {Huang}, {Zhang}, {Pei}, \&
  {Fan}}]{Ye_2023}
{Ye}, X.-H., {Zeng}, X.-T., {Huang}, D.-Y., {et~al.} 2023, \pasp, 135, 014101

\end{thebibliography}


\listofobjects 

\end{document}